\documentclass[twoside, journal]{IEEEtran}

\usepackage[utf8]{inputenc}
\usepackage{amsfonts}
\usepackage{amssymb,amsthm,epsfig}
\usepackage{amsmath}
\usepackage{graphicx,color,adjustbox}
\usepackage{dsfont,subfigure}
\usepackage{fixmath}
\graphicspath{ {figures/} }
\usepackage[linesnumbered,lined,boxed,commentsnumbered]{algorithm2e}
\usepackage{cite}
\usepackage{rotating}
\usepackage{longtable}
\usepackage{multicol}
\usepackage{boldline,multirow}

\SetAlCapSkip{0.5em}

\newtheorem{thm}{Theorem} 

\newtheorem{cor} {Corollary}
\newtheorem{Def} {Definition}

\newtheorem{examp}{Example} 
\newtheorem{asum}{Assumption}
\newtheorem{remark}{Remark}

\begin{document}
\title{A Generalized Unscented Transformation for Probability Distributions}
\author{Donald~Ebeigbe,
        Tyrus~Berry,
        Michael M. Norton, Andrew J. Whalen, Dan Simon, Timothy Sauer, and~Steven J. Schiff
\thanks{This work was supported by
NIH Director’s Transformative Award No. 1R01AI145057,
and from the National Science Foundation DMS-1723175,
DMS-1854204, and DMS-2006808. (\textit{Corresponding author: dee5127@psu.edu}).}

\thanks{D. Ebeigbe and M. M. Norton are with the Center for Neural Engineering, Department of Engineering Science and Mechanics, Pennsylvania State University, University Park, PA, USA (email: dee5127@psu.edu; mmn5439@psu.edu).}

\thanks{T. Berry and T. Sauer are with the Department of Mathematical Sciences, George Mason University, Fairfax, VA, USA (email: tberry@gmu.edu; tsauer@gmu.edu ).}

\thanks{A. J. Whalen is with the Center for Neural Engineering, Department of Engineering Science and Mechanics, Pennsylvania State University, University Park, PA, USA, and also with the Department of Neurosurgery, Massachusetts General Hospital, Harvard Medical School, Boston, MA, USA (email: awhalen7@mgh.harvard.edu  ).}

\thanks{D. Simon is with the Department of Electrical Engineering and Computer Science, Cleveland State University, Cleveland, OH, USA (email: d.j.simon@csuohio.edu).}

\thanks{S. J. Schiff is with the Center for Neural Engineering and Center for Infectious Disease Dynamics, Departments of Engineering Science and Mechanics, Neurosurgery, and Physics, Pennsylvania State University, University Park, PA, USA (email: sschiff@psu.edu).}
}


\maketitle

\begin{abstract}
The unscented transform uses a weighted set of samples called sigma points to propagate the means and covariances of nonlinear transformations of random variables. However, unscented transforms developed using either the Gaussian assumption or a minimum set of sigma points typically fall short when the random variable is not Gaussian distributed and the nonlinearities are substantial. In this paper, we develop the generalized unscented transform (GenUT), which uses $2n+1$ sigma points to accurately capture up to the diagonal components of the skewness and kurtosis tensors of most probability distributions. Constraints can be analytically enforced on the sigma points while guaranteeing at least second-order accuracy. The GenUT uses the same number of sigma points as the original unscented transform while also being applicable to non-Gaussian distributions, including the assimilation of  observations in the modeling of infectious diseases such as coronavirus (SARS-CoV-2) causing COVID-19.
\end{abstract}


\IEEEpeerreviewmaketitle

\section{Introduction}
\IEEEPARstart{T}{he} Kalman filter provides the basis for most of the popular state estimation techniques used for linear and nonlinear dynamic systems. The linear Kalman filter works by propagating the means and covariance of the state of a dynamic system~\cite{simon2006optimal, kandepu2008constrained}. Originally developed under the Gaussian assumption for measurement and process noise, the Kalman filter is the optimal estimator when this assumption is satisfied. Under non-Gaussian noise, the Kalman filter is the optimal linear estimator but its performance can sometimes deteriorate~\cite{simon2006optimal, izanloo2016kalman}. 

For many dynamic systems in practice, linearity is a reasonable assumption. For others, system non-linearities cause methods based on linear models to perform poorly. Most nonlinear systems can behave approximately linearly over small operation ranges. The extended Kalman filter (EKF) is one of the most widely used Kalman filter for nonlinear dynamic systems. The EKF employs a linear approximation of the nonlinear system around a nominal state trajectory~\cite{simon2006optimal, kandepu2008constrained, gustafsson2011some}. However, for highly nonlinear systems, linear approximations can introduce errors that can lead to divergence of the state estimate. 

To address the drawbacks of the EKF, several well-known state estimators such as the ensemble Kalman filter~\cite{EnKF1,EnKF2,anderson2001ensemble,berry2013adaptive}, the unscented Kalman filter (UKF)~\cite{julier1996general, julier2002reduced}, and the particle filter~\cite{simon2006optimal, kitagawa1996monte} have been developed. 
Although the particle filter can give better performance than the UKF, this comes at the cost of a higher computational effort. In some applications, the improved performance might not be worth the additional computational costs~\cite{simon2006optimal}. 

The UKF is a nonlinear filter that uses the unscented transformation to approximate the mean and covariance of a Gaussian random variable~\cite{julier1996general, julier1997consistent}. The unscented transform uses the intuition that \textit{with a fixed number of parameters it should be easier to approximate a Gaussian distribution than it is to approximate an arbitrary nonlinear function or transformation}~\cite{julier1996general}. It produces sets of vectors called \textit{sigma points} that capture the moments of the standard Gaussian distribution. The UKF uses the generated sigma points to obtain estimates of the states and the state estimation error covariance. The UKF has been used to generate distributions which improve the performance of a particle filter~\cite{rui2001better, van2001unscented}. It has also been employed to improve the performance of the EnKF~\cite{luo2009ensemble}. Despite the several types of sigma points that exist in the literature~\cite{simon2010kalman,cheng2011optimized}, a majority of them that were not developed using the Gaussian assumption do not try to match the skewness or kurtosis of a random variable, thereby ensuring only second-order accuracy.

The need to effectively monitor, predict, and control the spread of infectious disease has led to the application of numerous state estimation techniques. The EKF~\cite{simons2012assessment, chen2012tracking} and the particle filter~\cite{breto2009time} have been used to estimate the parameters of the measles virus transmission dynamics from real data. The ensemble adjustment Kalman filter (EAKF) has been employed in the forecasting of influenza~\cite{shaman2012forecasting} and dengue fever~\cite{yamana2016superensemble}. Several infectious disease such as Ebola~\cite{ndanguza2017analysis}, HIV~\cite{cazelles1997using}, and neonatal sepsis~\cite{ebeigbe2020poisson} have seen implementation of different Kalman filters. More recently, the outbreak of the novel coronavirus (SARS-CoV-2) causing COVID-19 has led to concerted efforts to properly understand its transmission and offer policy guidelines that can mitigate its spread. Recent efforts have employed the iterated EAKF to assimilate daily observations in the modeling of COVID-19~\cite{li2020substantial}. Distributions such as Poisson, negative-binomial, and binomial are typically used for modeling infectious disease from count data. Additionally, the number of patients arriving at a hospital or a testing center can be modeled by a Poisson distribution whose rate is proportional to the infected population. Although the use of standard Kalman filters in infectious disease estimation and prediction under the Poisson assumption can be justified with the fact that a Poisson distribution with a large rate can be approximated by a Gaussian distribution of the same mean and variance, the approximation breaks down when the rates are small~\cite{curtis1975simple}. 

The usage of Kalman filters to assimilate data generated by the transformation of random variables from different probability distributions revealed a fundamental mismatch in the application of the filters -- the accuracy of the filter is reduced if the Gaussian assumption is not satisfied and the nonlinearities are high. This led to the development of unscented transforms that can account for some higher-order moment information such as the skewness and kurtosis \cite{ponomareva2010new,straka2012randomized, rezaie2016skewed, hou2019high, easley2021higher}. The unscented transforms can be grouped into two categories: the ones that employ $2n +1$ sigma points \cite{ponomareva2010new, straka2012randomized, rezaie2016skewed} and the ones that use more than $2n+1$ sigma points \cite{hou2019high, easley2021higher}. 

First, we consider those that use $2n+1$ sigma points. In \cite{ponomareva2010new}, an unscented transform was developed to match the average marginal skewness and kurtosis. The method however did not match the true skewness and true kurtosis for each element of the random vector. In \cite{straka2012randomized}, a randomized unscented transform was used in the development of a filter for non-Gaussian systems. Although the method uses a stochastic integration rule to solve state and measurement statistics, the sigma points are generated under the Gaussian assumption.  In \cite{rezaie2016skewed}, an unscented transform was developed to capture the skewness of a random vector. However, the method assumes a closed skew normal distribution in its development. All preceding methods that use $2n+1$ sigma points either apply only to special distributions or can capture at most the average skewness and kurtosis.

Now we consider those that use more than $2n+1$ sigma points. In \cite{hou2019high}, an unscented transform was developed to match the first four moments of Gaussian random variables. In \cite{easley2021higher}, a higher order unscented transform was developed to match the skewness and kurtosis tensors with high accuracy. The method uses an approximate CANDECOMP/PARAFAC (CP) tensor decomposition to generate its sigma points. However, depending on the dimension of the problem and the error tolerance level in approximating the skewness and kurtosis tensors, this method can require significant computational costs. This is because the sequence of vectors and constants used in the approximate CP method can significantly increase when the error tolerance level is made small. All preceding methods that use more than $2n+1$ sigma points either applied to only to special distributions or had significantly higher complexity and computational cost.  

For an $n$-dimensional random vector, $2n +1$ sigma points generally employs $2n^2 + 3n + 1$ free parameters ($2n +1$ weights and $2n^2 +n$ constants that define the coordinates of the sigma points). Trying to match the mean, covariance, skewness, and kurtosis imposes $n$, $O(n^2)$, $O(n^3)$, and $O(n^4)$ constraints respectively. In principle, it is impossible to match all these moments using only $2n+1$ sigma points. The zero skewness nature of the Gaussian distribution made it possible to use $2n+1$ sigma points to accurately match up to the skewness in \cite{julier1996general}. The presence of the $O(n^3)$ skewness and $O(n^4)$ kurtosis constraints are what prompted researchers to look beyond $2n + 1$ sigma points. However, we note that matching the mean and covariance constraints of any random vector using $2n +1$ sigma points still leaves $n^2 + 2n +1$ free parameters. These residual parameters have been underutilized in capturing as much information as possible about the components of the skewness and kurtosis tensors when the random variable is not Gaussian. One instance where the residual parameters were leveraged was in the capturing of the average marginal skewness and kurtosis, which only represents a total of $2$ constraints \cite{ponomareva2010new}.

In this paper, we develop the \textit{generalized unscented transform} (GenUT) which is able to adapt to the unique statistics of most probability distributions. We use the intuition that \textit{employing sigma points more suitable to the inherent distributions of a random vector can lead to a more accurate propagation of means and covariances}. Our method uses $2n +1$ sigma points that not only accurately matches the mean and covariance matrix, but also takes advantage of the additional free parameters to accurately  match the diagonal components of the skewness tensor and kurtosis tensor of most random vectors. We employ $\frac{n^2 + n}{2} + 3n$ constraints in total; $n$ for the mean, $\frac{n^2 + n}{2}$ for the covariance, $n$ for the diagonal components of the skewness tensor, and $n$ for the diagonal components of the kurtosis tensor. This total falls within the $2n^2 + 3n +1$ free parameters available. While more parameters remain, the diagonal components of the skewness and kurtosis tensors are the most significant. In comparison to \cite{ponomareva2010new,straka2012randomized, rezaie2016skewed, hou2019high}, our method gives a general way to accurately match the diagonal components of the skewness and kurtosis tensors of most random vectors. In comparison to \cite{easley2021higher}, our method uses fewer sigma points which is crucial for larger system dimensions. In comparison to the standard unscented transform, we acquire the most significant higher moment information of most probability distributions with the same number of sigma points.

In Section \ref{section_gaussAssum}, we discuss the problems that arise when the Gaussian assumption is employed in the unscented transform. In Section~\ref{section_PUT}, we develop the GenUT sigma points that can capture certain properties of most probability distributions, such as its mean, covariance, skewness, and kurtosis. In Section~\ref{section_sigmaAccuracy}, we show that our sigma points are accurate in approximating the mean, covariance, and diagonal components of the skewness and kurtosis tensors. In Section~\ref{section_sigmaConstrained}, we address constraints and show that imposing constraints can at least maintain second-order accuracy. In Section~\ref{section_propagation}, we evaluate the accuracy of the GenUT sigma points in propagating means and covariances of nonlinear transformations of arbitrarily distributed random vectors and we give several examples that demonstrate its effectiveness when compared against other unscented transforms. We discuss the conclusions in Section~\ref{section_conclusion}.

\section{Limitations of the Unscented Transform}
\label{section_gaussAssum}
We analyze the performance of unscented transforms that were motivated by the Gaussian statistics \cite{julier1996general,julier1997consistent}.  We will show how linearization approximations, via Taylor series expansion of a nonlinear transformation of a random vector $\boldsymbol{x}$ evaluated about its mean $\bar{\boldsymbol{x} }$, introduces errors in the propagation of means and covariances. We will see that errors can be introduced in the propagation of means and covariances beyond the second order when used to approximate a nonlinear function $\boldsymbol{\lambda}(\boldsymbol{x})$ of a possibly non-Gaussian distributed random vector $\boldsymbol{x}\in \mathbb{R}^n$.
\begin{Def}
\label{def_xMoments}
Let $\boldsymbol{x} \in \mathbb{R}^n$ be a random vector. We define the mean $\bar{\boldsymbol{x}}\in \mathbb{R}^n$, covariance $\boldsymbol{P}\in \mathbb{R}^{n \times n}$, skewness tensor $\boldsymbol{S}\in \mathbb{R}^{n \times n \times n }$, and kurtosis tensor $\boldsymbol{K}\in \mathbb{R}^{n \times n \times n \times n}$ as  
%
\begin{align}
\bar{\boldsymbol{x}} &= \mathbb{E}[\boldsymbol{x}]  \\
\boldsymbol{P} &=  \mathbb{E} [ ( \boldsymbol{x} - \bar{\boldsymbol{x}}) ( \boldsymbol{x} - \bar{\boldsymbol{x}})^T ]\\
\boldsymbol{S}_{ijk} &= \mathbb{E}\left[ ( \boldsymbol{x} - \bar{\boldsymbol{x}})_i ( \boldsymbol{x} - \bar{\boldsymbol{x}})_j (  \boldsymbol{x}- \bar{\boldsymbol{x}})_k \right]   \label{eqn_skew_tens} \\
\boldsymbol{K}_{ijkl} &= \mathbb{E}\left[ (\boldsymbol{x} - \bar{\boldsymbol{x}})_i (\boldsymbol{x} - \bar{\boldsymbol{x}})_j (\boldsymbol{x} - \bar{\boldsymbol{x}})_k (\boldsymbol{x} - \bar{\boldsymbol{x}})_l \right]   \label{eqn_kurt_tens}
\end{align}
for $i,j,k,l \in \{1, \cdots,  n\}$.
\end{Def}

The sample mean and sample covariance of the nonlinear transformation $\boldsymbol{y} \in \mathbb{R}^n$ given by
\begin{align}
\boldsymbol{y}  =  \boldsymbol{\lambda}(\boldsymbol{x}) \label{eqn_YdynProb}
\end{align}
can be calculated as follows \cite{julier1996general}.
\begin{enumerate}
    \item Calculate the $2n+1$ sigma points  given by \footnote{Bold fonts are used to represent vectors, matrices, and tensors.}  
   \begin{align*}
   \boldsymbol{\chi}_{[0]} & = \bar{\boldsymbol{x}} \qquad \qquad \qquad \qquad \qquad \boldsymbol{w}_0  = \frac{\kappa}{n+\kappa}    \\
\boldsymbol{\chi}_{[i]} & = \bar{\boldsymbol{x}} + \left(\sqrt{(n + \kappa) \boldsymbol{P}}\right)_{[i]} \qquad   \boldsymbol{w}_i  = \frac{1}{2(n+\kappa)} \\
   \boldsymbol{\chi}_{[i+n]} & = \bar{\boldsymbol{x}} - \left(\sqrt{(n + \kappa)\boldsymbol{P}} \right)_{[i]}\qquad   \boldsymbol{w}_{i+n}  = \frac{1}{2(n+\kappa)}   
\end{align*}    
  for $i \in \{1, \cdots, n\}$, where $\left(\sqrt{(n + \kappa) \boldsymbol{P}}\right)_{[i]}$ is the $i$th column of $ \sqrt{(n + \kappa) \boldsymbol{P}}$, $\boldsymbol{w}_i$ is the weight associated with the $i$th sigma point, and $\kappa$ is a free parameter \footnote{The notation $\boldsymbol{P}_{[i]}$ represents the $i$th column of the matrix $\boldsymbol{P}$, $\boldsymbol{P}_{ij}$ represents the $i$th entry in the $j$th column of the matrix $\boldsymbol{P}$, and $\boldsymbol{x}_i$ represents the $i$th entry of the vector $\boldsymbol{x}$.} We typically set $\kappa = n-3$ to minimize the fourth-order moment mismatch.
    \item Pass the sigma points through the known nonlinear function to get the transformed sigma points
    \begin{align}
   \boldsymbol{\mathcal{Y}} _{[i]}  = \boldsymbol{\lambda}(\boldsymbol{\chi}_{[i]} )
   \label{eqn_tr_step2}
    \end{align}
    
    \item Evaluate the sample mean of the transformed sigma points
    \begin{align}
        \bar{\boldsymbol{y}} = \sum_{i = 0}^{2n} \boldsymbol{w}_i \boldsymbol{\mathcal{Y}} _{[i]}  \label{eqn_tr_step3}
    \end{align}
    
    \item Evaluate the sample covariance of the transformed sigma points
    \begin{align}
        \boldsymbol{P}_{y} =  \sum_{i = 0}^{2n} \boldsymbol{w}_i (\boldsymbol{\boldsymbol{\mathcal{Y}}}_{[i]}  - \bar{\boldsymbol{y}})(\boldsymbol{\boldsymbol{\mathcal{Y}}} _{[i]}  - \bar{\boldsymbol{y}})^T \label{eqn_tr_step4}
    \end{align}
\end{enumerate}

\subsection{Accuracy in Approximating the True Mean}
Applying a Taylor series expansion of $\boldsymbol{\lambda}(\boldsymbol{x})$ about its mean $\bar{\boldsymbol{x}}$, we show in Appendix \ref{Appendix_trueMean} that the true mean of $\boldsymbol{y} = \boldsymbol{\lambda}(\boldsymbol{x})$ is given as

\begin{align}
\bar{\boldsymbol{y}}  = &  \boldsymbol{\lambda}(\bar{\boldsymbol{x}}) +  \left\{     \sum_{i,j = 1}^{n} \frac{\boldsymbol{P}_{ij} }{2!}   \frac{\partial^2 \boldsymbol{\lambda}}{\partial  \boldsymbol{x}_i \partial  \boldsymbol{x}_j}       +    \sum_{i,j,k = 1}^{n} \frac{\boldsymbol{S}_{ijk} }{3!}   \frac{\partial^3 \boldsymbol{\lambda}}{\partial  \boldsymbol{x}_i \partial  \boldsymbol{x}_j \partial  \boldsymbol{x}_k }      \right. \nonumber \\
& \left. +   \sum_{i,j,k,l = 1}^{n}   \frac{\boldsymbol{K}_{ijkl}}{4!} \frac{\partial^4 \boldsymbol{\lambda}}{\partial  \boldsymbol{x}_i \partial  \boldsymbol{x}_j \partial  \boldsymbol{x}_k \partial  \boldsymbol{x}_l }     \right\}_{\boldsymbol{x} = \bar{\boldsymbol{x} }}  \nonumber \\
 & + \mathbb{E}\left[  \frac{D_{\tilde{x}}^5 \boldsymbol{\lambda}}{5!} + \frac{D_{\tilde{x}}^6 \boldsymbol{\lambda} }{6!}+ \cdots \right]  
 \label{eqn_trueMeanProb}
\end{align}
The analytical expression for the approximated mean from~\cite{julier1996general} is given as
\begin{align}
\bar{\boldsymbol{y}}_u &=  \boldsymbol{\lambda}(\bar{\boldsymbol{x}}) + \frac{1}{2} \sum_{i,j = 1}^{n}        \boldsymbol{P}_{ij}   \left. \frac{\partial \boldsymbol{\lambda} }{\partial  \boldsymbol{x}_i \partial  \boldsymbol{x}_j}       \right|_{\boldsymbol{x} = \bar{\boldsymbol{x} }}   \nonumber \\
& \quad + \frac{1}{2(n + \kappa)} \sum_{i = 1}^{2n} \left( \frac{D_{\sigma_i}^4 \boldsymbol{\lambda}}{4!} + \frac{D_{\sigma_i}^6 \boldsymbol{\lambda}}{6!}    + \cdots \right)
\end{align}
Comparing the above equation with the true mean of~(\ref{eqn_trueMeanProb}), we notice the following problems about the sigma points developed using the Gaussian assumption
\begin{enumerate}
\item The odd-powered moments in the approximation of the true mean are always zero due to their symmetry. This introduces significant approximation errors in situations where the odd-powered moments of the distribution of $\boldsymbol{x}$ are non-zero and the transformation $\boldsymbol{y} = \boldsymbol{\lambda}(\boldsymbol{x})$ is highly nonlinear.

\item The fourth-order term fails to capture a part of the true kurtosis even when the optimal value of $\kappa = n-3$ is selected because of the Gaussian assumption.

\end{enumerate}
We also note that errors in approximating the mean beyond the second order occur not only for sets of $2n +1$ sigma points existing in the literature, but also for sets of $n+1$ sigma points \cite{julier2002reduced, menegaz2015systematization} -- this is because they do not account for the skewness and kurtosis of $\boldsymbol{x}$ when it is not Gaussian distributed. 

\subsection{Accuracy in Approximating the True Covariance Matrix}
The true covariance matrix, which was evaluated in Appendix \ref{Appendix_covarianceTrue}, is given as
\begin{align}
P_y = & \boldsymbol{\lambda} \boldsymbol{P}  \boldsymbol{\lambda}^T  +  
 \left\{    \sum_{i,j,k  = 1}^{n} \frac{\boldsymbol{S}_{ijk}}{2!}   \left[   \frac{\partial^2 \boldsymbol{\lambda}}{\partial  \boldsymbol{x}_i \partial  \boldsymbol{x}_j}  \frac{\partial \boldsymbol{\lambda}^T}{\partial  \boldsymbol{x}_k}   +    \frac{\partial \boldsymbol{\lambda}}{\partial  \boldsymbol{x}_i}    \frac{\partial^2 \boldsymbol{\lambda}^T}{\partial  \boldsymbol{x}_j \partial  \boldsymbol{x}_k}  \right]  \right.
 \nonumber \\
& +  \sum_{i,j,k,l = 1}^{n}  \boldsymbol{K}_{ijkl}     \left[ \frac{1}{3!} \frac{\partial^3 \boldsymbol{\lambda}}{\partial  \boldsymbol{x}_i \partial  \boldsymbol{x}_j \partial  \boldsymbol{x}_k}  \frac{\partial \boldsymbol{\lambda}^T}{\partial  \boldsymbol{x}_l}  \right.  \nonumber \\
 & \left. +  \frac{1}{3!} \frac{\partial \boldsymbol{\lambda}}{\partial  \boldsymbol{x}_i} \frac{\partial^3 \boldsymbol{\lambda}^T}{\partial  \boldsymbol{x}_j \partial  \boldsymbol{x}_k \partial  \boldsymbol{x}_l} + \frac{1}{4} \frac{\partial^2 \boldsymbol{\lambda}}{\partial  \boldsymbol{x}_i \partial  \boldsymbol{x}_j  }  \frac{\partial^2 \boldsymbol{\lambda}^T}{\partial  \boldsymbol{x}_k \partial  \boldsymbol{x}_l  }   \right]   \nonumber \\
& + 
 \left. \left[     \left. \left.  \sum_{i,j = 1}^{n} \frac{\boldsymbol{P}_{ij}}{2}  \frac{\partial^2 \boldsymbol{\lambda}}{\partial  \boldsymbol{x}_i \partial  \boldsymbol{x}_j}      \right] \right[  \cdots  \right]^T   \right\}_{\boldsymbol{x} = \bar{\boldsymbol{x} }} + \cdots  
 \label{eqn_trueCovProb}
\end{align}
where we have used the notation $\boldsymbol{x} \boldsymbol{x}^T = \boldsymbol{x} [\cdots]^T$. The analytical expression for the approximated covariance matrix from~\cite{julier1996general} is given as 
\begin{align}
 \boldsymbol{P}_u = & \boldsymbol{\lambda} P \boldsymbol{\lambda}^T  +  \frac{1}{2(n + \kappa)} \sum_i^{2n} \left(\frac{  D_{\sigma_i} \boldsymbol{\lambda} (D_{\sigma_i}^3 \boldsymbol{\lambda})^T}{3!} \right. \nonumber \\
 & \left. + \frac{ D_{\sigma_i}^3 \boldsymbol{\lambda} ( D_{\sigma_i} \boldsymbol{\lambda})^T}{3!}  + \frac{ D_{\sigma_i}^2 \boldsymbol{\lambda} ( D_{\sigma_i}^2 \boldsymbol{\lambda})^T}{2! \times 2!} \right)
\nonumber \\
&   +  \left. \left.  \left[ \frac{1}{2}     \sum_{i,j = 1}^{n}  \boldsymbol{P}_{ij} \left.\frac{\partial^2 \boldsymbol{\lambda}}{\partial  \boldsymbol{x}_i \partial  \boldsymbol{x}_j}   \right|_{\boldsymbol{x} = \bar{\boldsymbol{x} }} \right] \right[   \cdots \right] ^T + \cdots   \label{eqn_coxApp2full}
\end{align}
Comparing the above equation with the true covariance matrix of~(\ref{eqn_trueCovProb}), we notice similar issues that were pointed out in approximating the mean -- the approximation is only accurate up to the second order when $\boldsymbol{x}$ is not Gaussian distributed. All the odd-powered moments are zero because of the symmetric nature of the sigma points, while the fourth-powered moment is also inaccurate because of the Gaussian nature of the sigma points. As with the mean approximation, errors in the covariance matrix approximation are introduced beyond the second order not only for sets of $2n +1$ sigma points existing in the literature, but also for sets of $n+1$ sigma points.  


\section{Generalized Unscented Transform}
\label{section_PUT}
For a random vector $\boldsymbol{x} \in \mathbb{R}^n$, we develop sigma points that can accurately capture the mean, covariance matrix, and the diagonal components of both the skewness tensor and the kurtosis tensor. This is done by selecting sigma point distributions that have the flexibility to either be symmetric when $\boldsymbol{x}$ is symmetrically distributed or be asymmetric when $\boldsymbol{x}$ is asymmetrically distributed. 

\begin{asum}
\label{Asum_1}
The random vector $\boldsymbol{x}$ follows a probability distribution with finite moments.
\end{asum}

We reduce the problem of approximating $\boldsymbol{x}$ to the problem of approximating a user-specified arbitrarily distributed random vector $\boldsymbol{z}\in \mathbb{R}^{n}$ with zero mean and unit variance, whose higher-order moments are functions of the higher-order moments of $\boldsymbol{x}$. We write 
\begin{align}
    \boldsymbol{x} = \bar{\boldsymbol{x}} +  \left(\sqrt{\boldsymbol{P}} \right) \boldsymbol{z} \label{eqn_Xmain}
\end{align}
where $\sqrt{\boldsymbol{P}}$ is the matrix square root of $\boldsymbol{P}$, $\sqrt{\boldsymbol{P}} \sqrt{\boldsymbol{P}}^T = \boldsymbol{P}$

\begin{Def}
\label{def_prodExplain}
Let $\boldsymbol{x}$ be a vector, $\boldsymbol{P}$ be a square matrix, and $k$ be some positive integer. We define the element-wise product (Hadamard product) as $\odot$, such that
\begin{align*}
     \boldsymbol{x}^{\odot k} = \underbrace{ \boldsymbol{x} \odot \boldsymbol{x} \odot \cdots   \odot \boldsymbol{x}}_\text{k times}  \\
     \boldsymbol{P}^{\odot -k} =  \left(\underbrace{ \boldsymbol{P} \odot \boldsymbol{P} \odot \cdots   \odot \boldsymbol{P}}_\text{k times}  \right)^{-1}
\end{align*}
We also define the element-wise division (Hadamard division) as $\oslash$.
\end{Def}

\subsection{One-Dimensional Distribution}
\begin{figure}
\centering     
\includegraphics[scale = 0.7]{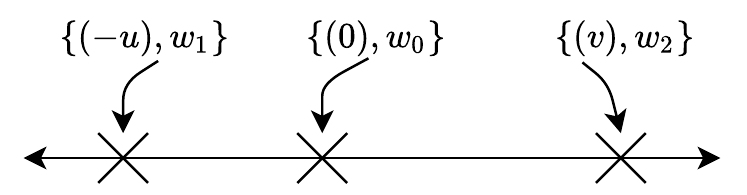}
\caption{  Samples chosen for a one-dimensional distribution for the GenUT. The locations and weights of the sigma points are determined by the moments of the probability distribution.}
\label{fig_oneDimension}
\end{figure}
%
We develop sigma points that match the first three moments of $\boldsymbol{z}$ in a single dimension, and then constrain those points to match the fourth moment of $\boldsymbol{z}$. For a one-dimensional distribution, we will show how to select sigma points such that the first four moments satisfy
\begin{align*}
\mathbb{E}\left[ \boldsymbol{z}_i \right] & = 0,    \qquad \quad \:\:\:\:
\mathbb{E} \left[ (\boldsymbol{z} - \bar{\boldsymbol{z}} )^2 \right] = 1     \\
\mathbb{E}\left[ (\boldsymbol{z} - \bar{\boldsymbol{z}} )^3 \right]  & = \frac{\boldsymbol{S} }{\sqrt{\boldsymbol{P}}^3}, \qquad
\mathbb{E}\left[ (\boldsymbol{z} - \bar{\boldsymbol{z}} )^4 \right] =  \frac{\boldsymbol{K} }{ \boldsymbol{P}^2}  
\end{align*}

To capture the first three moments in a single dimension, three points are used: the first point lies at the origin with a weight of $\boldsymbol{w}_0$; the second point lies at a distance $-\boldsymbol{u}$ from the origin with a weight of $\boldsymbol{w}_1$; the third point lies at a distance $\boldsymbol{v}$ from the origin with a weight of $\boldsymbol{w}_2$. Therefore, in one-dimension, we use the following 3 sigma points
\begin{align}
\boldsymbol{\chi}^{(0)} & = \{0\}, \boldsymbol{w}_0 \nonumber \\
\boldsymbol{\chi}^{(1)} & = \{-\boldsymbol{u}  \}, \boldsymbol{w}_1    \nonumber \\
\boldsymbol{\chi}^{(2)} & = \{\boldsymbol{v}  \},\boldsymbol{w}_2 \nonumber   
\end{align}
where $\boldsymbol{w}_0$, $\boldsymbol{w}_1$, and $\boldsymbol{w}_2$ are the weights for the respective sigma points. A visual representation of our sigma points in one dimension is shown in Fig.~\ref{fig_oneDimension}. Obeying the moments of $\boldsymbol{z}$ and the fact that the sum of all weights should equal 1, we write
\begin{align}
\boldsymbol{\boldsymbol{w}_0} + \boldsymbol{w}_1 + \boldsymbol{w}_2 & = 1  \label{eqn_sig_weight}  \\
-\boldsymbol{w}_1 \boldsymbol{u} + \boldsymbol{w}_2 \boldsymbol{v} & = 0   \label{eqn_sig_Mean} \\
 \boldsymbol{\boldsymbol{w}_1} \boldsymbol{\boldsymbol{u}}^2 + \boldsymbol{w}_2 \boldsymbol{v}^2 & = 1  \label{eqn_sig_Cov}  \\
  -\boldsymbol{w}_1 \boldsymbol{u}^3 + \boldsymbol{w}_2 \boldsymbol{v}^3 & =  \boldsymbol{S}  \sqrt{\boldsymbol{P}}^{-3}   \label{eqn_sig_Skew}
\end{align}
From (\ref{eqn_sig_Mean}), we see that $
\boldsymbol{\boldsymbol{w}_1} = \frac{\boldsymbol{v}}{\boldsymbol{u}}  \boldsymbol{w}_2 $.
Rewriting~(\ref{eqn_sig_Cov}) using~(\ref{eqn_sig_Skew}) gives
\begin{align}
    \boldsymbol{w}_2 \boldsymbol{v} (\boldsymbol{u} + \boldsymbol{v}) = 1  \label{eqn_wSub1}\\
    \boldsymbol{w}_2 \boldsymbol{v} ( \boldsymbol{v}^2 - \boldsymbol{u}^2) = \boldsymbol{S}  \sqrt{\boldsymbol{P}}^{-3} \label{eqn_wSub2}
\end{align}
We designate $\boldsymbol{u}$ as the free parameter while assuming that $\boldsymbol{u} > 0$. Using the fact that $\boldsymbol{v}^2 - \boldsymbol{u}^2 = (\boldsymbol{u} + \boldsymbol{v})(\boldsymbol{v} - \boldsymbol{u})$, substituting (\ref{eqn_wSub1}) into (\ref{eqn_wSub2}) gives
\begin{align}
    \boldsymbol{v}  = \boldsymbol{u} + \boldsymbol{S}  \sqrt{\boldsymbol{P}}^{-3} \label{eqn_1Dim_s2}
\end{align}
From~(\ref{eqn_sig_weight}) and (\ref{eqn_wSub1}), we see that the weights are given as
\begin{align}
   \boldsymbol{w}_2 = \frac{1}{\boldsymbol{v} (\boldsymbol{u} + \boldsymbol{v})}, \quad   \quad \boldsymbol{w}_0 = 1 - \boldsymbol{w}_1 - \boldsymbol{w}_2 \label{eqn_wghtsDim1}
\end{align}
We note that the free parameter $\boldsymbol{u}$ can be selected to match the fourth moment of $\boldsymbol{z}$. We now attempt to satisfy the fourth moment constraint given by 
\begin{align}
\boldsymbol{w}_1 \boldsymbol{u}^4 + \boldsymbol{w}_2 \boldsymbol{v}^4  = \boldsymbol{K}  \boldsymbol{P}^{-2}
\end{align}
 Eliminating $\boldsymbol{w}_1$ using $\boldsymbol{w}_1 = \frac{\boldsymbol{v}}{\boldsymbol{u}}  \boldsymbol{w}_2$ gives
\begin{align}
   \boldsymbol{w}_2 \boldsymbol{v} (\boldsymbol{u}^3 + \boldsymbol{v}^3) = \boldsymbol{K}  \boldsymbol{P}^{-2}  \label{eqn_wSub3}
\end{align}
Using the relationships $\boldsymbol{w}_2 \boldsymbol{v} (\boldsymbol{u} + \boldsymbol{v}) = 1 $, $\boldsymbol{u}^3 + \boldsymbol{v}^3 = (\boldsymbol{u} + \boldsymbol{v})(\boldsymbol{u}^2 + \boldsymbol{v}^2 - \boldsymbol{u} \boldsymbol{v})$, and $ \boldsymbol{v}  = \boldsymbol{u} + \boldsymbol{S}  \sqrt{\boldsymbol{P}}^{-3} $, the above equation reduces to
$$ \boldsymbol{u}^2 + \boldsymbol{S}  \sqrt{\boldsymbol{P}}^{-3} \boldsymbol{u} + \boldsymbol{S}^2  \boldsymbol{P}^{-3} - \boldsymbol{K}  \boldsymbol{P}^{-2} = 0  $$
The solution to the above quadratic equation is
\begin{align}
  \boldsymbol{u} = \frac{1}{2} \left[ - \boldsymbol{S}  \sqrt{\boldsymbol{P}}^{-3} + \sqrt{4\boldsymbol{K}  \boldsymbol{P}^{-2} - 3 \boldsymbol{S}^2  \boldsymbol{P}^{-3}   }   \right]
\end{align}
where $\boldsymbol{v}$ is given in~(\ref{eqn_1Dim_s2}). The equations for $\boldsymbol{w}_1$, $\boldsymbol{w}_2$, and $\boldsymbol{w}_0$ remain unchanged.

\begin{remark}
\label{remark_1Dim}
We note that the sigma points described above, which accurately capture the kurtosis when constrained, were designed for when the state has a dimension of 1. This implies that $\boldsymbol{z}, \boldsymbol{P}, \boldsymbol{S}, \boldsymbol{K} \in \mathbb{R}^1$.
\end{remark}
In the next section, we extend this to multiple dimensions.

\subsection{Multi-Dimensional Distribution}
For an $n$-dimensional vector $\boldsymbol{z}$, we develop a  set of sigma points that accurately matches its mean and covariance matrix, while accurately matching the diagonal components of the skewness tensor. Furthermore, by constraining the sigma points, we show that we can accurately match the diagonal components of the kurtosis tensor. We note that for an independent random vector, accurately matching the diagonal components of the skewness tensor implies an accurate matching of the entire skewness tensor.

\begin{Def}
\label{def_tensDiag}
We define the vectors $\breve{\boldsymbol{S}} \in \mathbb{R}^n$ and $\breve{\boldsymbol{K}} \in \mathbb{R}^n$ which contain the diagonal components of the skewness tensor and kurtosis tensor respectively, such that
\begin{align*}
    \breve{\boldsymbol{S}} &=  [\begin{matrix} \boldsymbol{S}_{111}, \boldsymbol{S}_{222}, \cdots, \boldsymbol{S}_{nnn}  \end{matrix}  ]^T \\
    \breve{\boldsymbol{K}} &=  [\begin{matrix} \boldsymbol{K}_{1111}, \boldsymbol{K}_{2222}, \cdots, \boldsymbol{K}_{nnnn}  \end{matrix}  ]^T
\end{align*}
\end{Def}
%
For a multi-dimensional distribution, we will show how to select the $2n+1$sigma points such that the first four moments satisfy 
\begin{align*}
\mathbb{E}\left[ \boldsymbol{z} \right] & = \boldsymbol{0} \\
\mathbb{E} \left[ (\boldsymbol{z} - \bar{\boldsymbol{z}} )  (\boldsymbol{z} - \bar{\boldsymbol{z}} )^T \right] &= \boldsymbol{I}     \\
\mathbb{E}\left[ (\boldsymbol{z} - \bar{\boldsymbol{z}} )^{\odot3} \right]  & =  \sqrt{\boldsymbol{P}}^{\odot -3}  \breve{\boldsymbol{S}}\\
\mathbb{E}\left[ (\boldsymbol{z} - \bar{\boldsymbol{z}} )^{\odot4} \right] &=   \sqrt{\boldsymbol{P}}^{\odot -4}  \breve{\boldsymbol{K}} 
\end{align*}
where $\boldsymbol{I} \in \mathbb{R}^{n \times n}$ is the identity matrix.

\begin{remark}
\label{remarkcovInv}
Due to the positive definiteness of the covariance matrix $\boldsymbol{P} \in \mathbb{R}^{n \times n}$, it is always invertible.
\end{remark}

\begin{figure}
\centering     
\includegraphics[scale = 0.7]{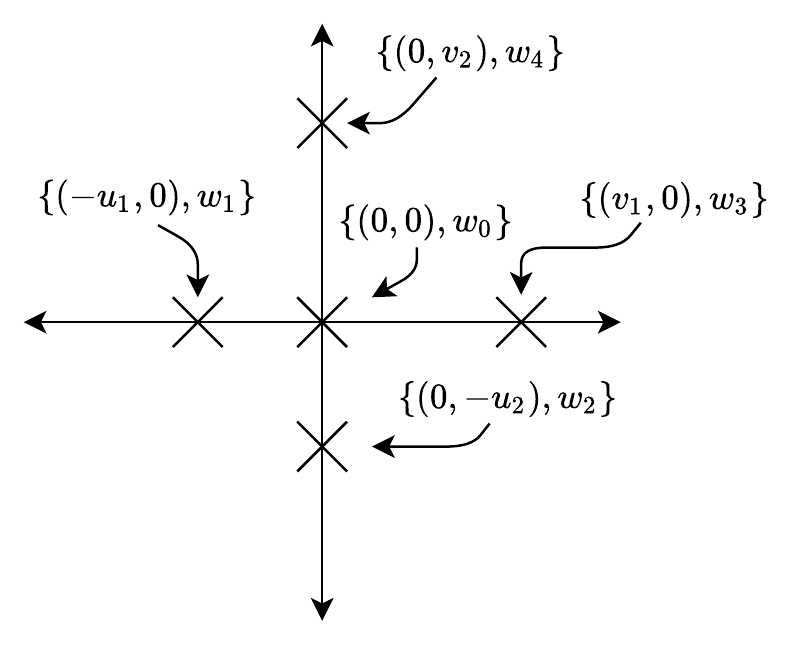}  
\caption{  Samples chosen for a two-dimensional distribution for the GenUT. The locations and weights of the sigma points are determined by the moments of the probability distribution.}
\label{fig_secondOrder}
\end{figure}

A visual representation of our sigma points for a two-dimensional distribution is shown in Fig.~\ref{fig_secondOrder}. Our first point lies at $(0,0)$ with a weight of $\boldsymbol{w}_0$. Our second point lies on the coordinate axes a distance $-\boldsymbol{u}_1$ from the origin with a weight of $\boldsymbol{w}_1$. Our third point lies on the coordinate axes a distance $-\boldsymbol{u}_2$ from the origin with a weight of $\boldsymbol{w}_2$. Our fourth point lies on the coordinate axes a distance $\boldsymbol{v}_1$ from the origin with a weight of $\boldsymbol{w}_3$. Our fifth point lies on the coordinate axes a distance $\boldsymbol{v}_2$ from the origin with a weight of $\boldsymbol{w}_4$. Therefore, our unscented transform uses the following $2n + 1$ sigma points
\begin{align}
\boldsymbol{\chi}^{(0)} & = \{\textbf{0}\}, \boldsymbol{w}_0 \nonumber \\
\boldsymbol{\chi}_{[i]} & = \{-\boldsymbol{u}_i \boldsymbol{I}_{[i]} \}, \boldsymbol{w}_i  \quad \qquad \: i = 1, \cdots, n   \nonumber \\
\boldsymbol{\chi}^{(i+n)} & = \{\boldsymbol{v}_{i}  \boldsymbol{I}_{[i]}\},\boldsymbol{w}_{i+n}  \qquad i = 1, \cdots, n  \nonumber
\end{align}
where $\boldsymbol{I}_{[i]}$ is the $i$th column of the identity matrix. $\textbf{0} \in \mathbb{R}^n$ is a vector of zeros. We note that $\boldsymbol{u} = [\begin{matrix} \boldsymbol{u}_1, \boldsymbol{u}_2, \cdots, \boldsymbol{u}_n  \end{matrix}]^T$ and $\boldsymbol{v} = [\begin{matrix} \boldsymbol{v}_1, \boldsymbol{v}_2, \cdots, \boldsymbol{v}_n  \end{matrix}]^T$ 
\begin{Def}
\label{def_vecSimpl}
We partition the weight vector $\boldsymbol{w} = [\begin{matrix} \boldsymbol{w}_0, \boldsymbol{w}_1, \cdots, \boldsymbol{w}_{2n}  \end{matrix}]^T$ by defining $\boldsymbol{w}' = [\begin{matrix} \boldsymbol{w}_1, \boldsymbol{w}_2, \cdots, \boldsymbol{w}_{n}  \end{matrix}]^T$ and $\boldsymbol{w}'' = [\begin{matrix} \boldsymbol{w}_{1+n}, \boldsymbol{w}_{2+n}, \cdots, \boldsymbol{w}_{2n}  \end{matrix}]^T$  such that $\boldsymbol{w} = [\begin{matrix} \boldsymbol{w}_0, \boldsymbol{w}'^T,  \boldsymbol{w}''^T  \end{matrix}]^T$. 
\end{Def}

Obeying the moments of $\boldsymbol{z}$, we write
\begin{align}
 \sum_{i = 0}^{2n} \boldsymbol{w}_i & = 1  \label{eqn_nDim_weight1}  \\
  -\boldsymbol{w}' \odot \boldsymbol{u} + \boldsymbol{w}'' \odot \boldsymbol{v} & = \boldsymbol{0}    \label{eqn_nDim_Mean}\\
  \boldsymbol{w}' \odot \boldsymbol{u}^{\odot 2} + \boldsymbol{w}'' \odot \boldsymbol{v}^{\odot 2} & = \boldsymbol{1} \label{eqn_nDim_Cov} \\
  -\boldsymbol{w}' \odot \boldsymbol{u}^{\odot 3} + \boldsymbol{w}'' \odot \boldsymbol{v}^{\odot 3} & =  \sqrt{\boldsymbol{P}}^{\odot -3}  \breve{\boldsymbol{S}} \label{eqn_nDim_Skew} 
\end{align}
where $\textbf{1} \in \mathbb{R}^n$ is a vector of ones. From~(\ref{eqn_nDim_Mean}), we see that $\boldsymbol{w}' =  \boldsymbol{w}'' \odot \boldsymbol{v} \oslash \boldsymbol{u} $. Rewriting~(\ref{eqn_nDim_Cov}) and (\ref{eqn_nDim_Skew}) gives
\begin{align}
\boldsymbol{w}'' \odot \boldsymbol{v} \odot ( \boldsymbol{u} +   \boldsymbol{v} ) & = \boldsymbol{1}    \label{eqn_sig_Cov1a} \\
  -\boldsymbol{w}'' \odot \boldsymbol{v} \odot ( \boldsymbol{u} +   \boldsymbol{v} ) \odot ( \boldsymbol{v} +   \boldsymbol{u} ) & =  \sqrt{\boldsymbol{P}}^{\odot -3}  \breve{\boldsymbol{S}} \label{eqn_sig_Skew1a} 
\end{align}
Selecting $\boldsymbol{u} > \boldsymbol{0}$ as the free parameters, we get
\begin{align}
    \boldsymbol{v}  = \boldsymbol{u} +  \sqrt{\boldsymbol{P}}^{\odot -3}  \breve{\boldsymbol{S}} \label{eqn_ndim_ssi}
\end{align}
Therefore, from~(\ref{eqn_nDim_weight1}) and (\ref{eqn_sig_Cov1a}), we see that 
\begin{align}
   \boldsymbol{w}'' =  \boldsymbol{1} \oslash \boldsymbol{v} \oslash ( \boldsymbol{u} +   \boldsymbol{v} ) , \quad
     \boldsymbol{w}_0 = 1 - \sum_{i = 1}^{2n} \boldsymbol{w}_i \label{eqn_2Dimw0}
\end{align}
To match the diagonal components of the kurtosis tensor, we need to satisfy
\begin{align}
  \boldsymbol{w}' \odot \boldsymbol{u}^{\odot 4} + \boldsymbol{w}'' \odot \boldsymbol{v}^{\odot 4} & =  \sqrt{\boldsymbol{P}}^{\odot -4}  \breve{\boldsymbol{K}}  \label{eqn_nDim_kurt}  
\end{align}
Solving the above equation results in constrained values for $\boldsymbol{u}$, such that
\begin{align}
  \boldsymbol{u} = \frac{1}{2} \left( -    \sqrt{\boldsymbol{P}}^{\odot -3} \breve{\boldsymbol{S}}   + \sqrt{ 4 \sqrt{\boldsymbol{P}}^{\odot -4}  \breve{\boldsymbol{K}}  - 3 \left(\sqrt{\boldsymbol{P}}^{\odot -3}  \breve{\boldsymbol{S}} \right)^{\odot 2} }  \right)    
  \label{eqn_sConstrained}
\end{align}
It can be shown from (\ref{eqn_Xmain}) that the algorithm  for selecting the $2n+1$ sigma points for any random vector $\boldsymbol{x}$ is given in Algorithm~\ref{algo_disjdecomp}. 

We recall from (\ref{eqn_ndim_ssi}) the constraint $\boldsymbol{u} > \boldsymbol{0}$ exists. Applying this constraint on (\ref{eqn_sConstrained}), we see that
\begin{align}
 \breve{\boldsymbol{K}}  & > \sqrt{\boldsymbol{P}}^{\odot 4}    \left(\sqrt{\boldsymbol{P}}^{\odot -3}   \breve{\boldsymbol{S}} \right)^{\odot 2} \label{eqn_skewKurt}
\end{align}
The inequality in (\ref{eqn_skewKurt}) -- at least for a one-dimensional case -- agrees with the findings by Pearson in \cite{pearson1916ix} that for probability distributions, the standardized kurtosis always exceeds the squared of the standardized skewness. If the inequality in (\ref{eqn_skewKurt}) were violated, then (\ref{eqn_sConstrained}) becomes infeasible, which in turn requires the free parameter $\boldsymbol{u} > \boldsymbol{0}$ in (\ref{eqn_ndim_ssi}) to be selected such that $\boldsymbol{v} > \boldsymbol{0}$ -- although this eliminates the accuracy in matching the diagonal components of the kurtosis tensor, the sigma points are still able to accurately match the diagonal components of the skewness tensor.

There might be concerns that $\boldsymbol{v}$ in (\ref{eqn_ndim_ssi}) might be negative whenever the term $   \sqrt{\boldsymbol{P}}^{\odot -3}  \breve{\boldsymbol{S}}$ is negative. If (\ref{eqn_skewKurt}) is satisfied, then selecting $\boldsymbol{u}$ using (\ref{eqn_sConstrained}) leads to $\boldsymbol{v} > \boldsymbol{0}$. Alternatively, arbitrarily selecting $\boldsymbol{u}$ such that $\boldsymbol{u} > \sqrt{\boldsymbol{P}}^{\odot -3} \breve{\boldsymbol{S}}$ ensures that $\boldsymbol{v} > \boldsymbol{0}$.

Algorithm~\ref{algo_disjdecomp} can be used to create sigma points that can match up to the kurtosis if (\ref{eqn_skewKurt}) is satisfied. For example, we want to prescribe some arbitrary mean, variance, skewness, and kurtosis for a random variable $x$ that is not from any known probability distribution. Randomly selecting the mean $\bar{\boldsymbol{x} }$, variance $\boldsymbol{P}$, and skewness $\boldsymbol{S}$ as $\bar{\boldsymbol{x} } = 0.1$, $\boldsymbol{P} = 0.2$, and $\boldsymbol{S} = -0.5$ respectively, we can use Algorithm~\ref{algo_disjdecomp} to match them exactly. However, we can not randomly select a kurtosis $\boldsymbol{K}$ and expect to match it. The selection of the kurtosis $\boldsymbol{K}$ must satisfy (\ref{eqn_skewKurt}), so for this example, we require $\boldsymbol{K} >  \frac{-0.5^2}{ 0.2} = 1.25$. Prescribing a kurtosis value of $\boldsymbol{K} = 1.3$ satisfies (\ref{eqn_skewKurt}). Now using Algorithm~\ref{algo_disjdecomp}, we see that $\boldsymbol{w}_0 = 0.2$, $\boldsymbol{w}_1 = 0.0286$, $\boldsymbol{w}_2 = 0.7714$, $\boldsymbol{u} = 5.8055 $, and $\boldsymbol{v} = 0.2153$. The sample mean, sample covariance, sample skewness, and sample kurtosis exactly match their true prescribed values. We show how to calculate the sample statistics in Section~\ref{section_sigmaAccuracy}.

\subsection{Moments of a Probability Distribution}
 We use the moment generating function (MGF) $M(t)$ to evaluate the mean and higher-order central moments of a probability distribution. For any random variable $\boldsymbol{x}$~\cite{papoulis2002probability}, its MGF and $n$-th moment are given by
\begin{align}
M(t) = \mathbb{E}[e^{t\boldsymbol{x}}]  \label{eqn_MGF}, \quad \mathbb{E}[\boldsymbol{x}^n] =  \left. \frac{\partial^n M}{\partial t^n} \right|_{t = 0}
\end{align}
We also use the gamma notation 
\begin{align}
\Gamma(k) = \int_0^\infty \boldsymbol{x}^{k-1} e^{-\boldsymbol{x}} \: dx
\end{align}
The first four moments of 10 different probability distributions can be found in Table~\ref{tab_probDistMoment}.

\begin{table*}
\small
\caption{Probability Distributions} 
\centering 
\begin{tabular}{l c c ccc } 
\hline\hline 
   & Probability density  & Mean & Variance $(\boldsymbol{P})$ & Skewness $(\boldsymbol{S})$ & Kurtosis $ (\boldsymbol{K})$    \\
 Random Variable &   function & $\mathbb{E}[\boldsymbol{x}]$  &$\mathbb{E}[ (\boldsymbol{x} - \bar{\boldsymbol{x}})^2]$ &  $\mathbb{E}[ (\boldsymbol{x} - \bar{\boldsymbol{x}})^3]$ & $\mathbb{E}[ (\boldsymbol{x} - \bar{\boldsymbol{x}})^4]$ 
\\   
\hline \hline 

{\rule{0pt}{7.5ex}} 
  Gaussian $\mathcal{N}(\mu, \sigma^2)$  & $\begin{aligned}[c] 
& \frac{1}{\sqrt{2 \pi \sigma^2}} e^{- \frac{1}{2} \left(\frac{x - \mu}{\sigma }\right)^2},\\
&x \in (-\infty,\: \infty) 
\end{aligned}$  & $\mu$ & $\sigma^2 $ & 0 & $ 3 \sigma^4 $    {\rule[-6ex]{0pt}{0pt}}\\ \hline

{\rule{0pt}{4ex}} 
Exponential $E(\lambda)$  & $ \lambda e^{-\lambda x}, x\geq 0, \lambda> 0 $  & $\frac{1}{\lambda }$ & $ \frac{1}{\lambda^2}  $ & $\frac{2}{\lambda^3}  $  & $\frac{9}{\lambda^4}$   {\rule[-4ex]{0pt}{0pt}} \\ \hline

{\rule{0pt}{7.5ex}} 
Gamma $G(a,b)$  & $\begin{aligned}[c] 
& \frac{x^{a-1}}{\Gamma(a) b^{a}} e^{\frac{-x}{b}} ,\\
& x\geq 0, a> 0, b>0
\end{aligned}$  & $a b $  & $a b^2 $  &  $ 2 a b^3 $ & $3a b^4 (a + 2)  $     {\rule[-6ex]{0pt}{0pt}} \\ \hline

{\rule{0pt}{17ex}} 
Weibull $W(a, b)$  & $\begin{aligned}[c] 
&\frac{b}{a} \left(\frac{x}{a} \right)^{b-1} e^{-\left(\frac{x}{a} \right)^b} ,\\
& x \geq 0, a> 0, b>0,\\
&\Gamma_{kb} = \Gamma\left(\frac{k}{b} +1  \right)
\end{aligned}$  & $a \Gamma_{1b}$ & $a^2 \left[ \Gamma_{2b}    -   \Gamma_{1b}^2   \right]   $  & $\begin{aligned}[c] 
&a^3 ( \Gamma_{3b}    + 2   \Gamma_{1b}^3 \\
& - 3\Gamma_{1b} \Gamma_{2b}   )
\end{aligned}$   & $\begin{aligned}[c] 
&a^4 ( \Gamma_{4b}    -3   \Gamma_{1b}^4  \\
& - 4\Gamma_{1b} \Gamma_{3b} \\ & + 6   \Gamma_{1b}^2 \Gamma_2)
\end{aligned}$    {\rule[-11ex]{0pt}{0pt}} \\ \hline

{\rule{0pt}{6ex}} 
Rayleigh $R(\sigma)$  & $\begin{aligned}[c] 
& \frac{x}{\sigma^2} e^{-\frac{x^2}{2\sigma^2} }, x \geq 0   
\end{aligned}$  &  $\sigma \sqrt{\frac{\pi}{2}}$  & $\sigma^2 \left(2 - \frac{\pi}{2} \right) $ &  $\sigma^3 \left(\pi - 3 \right) \sqrt{\frac{\pi}{2}}    $ & $\sigma^4 \left(\frac{32 - 3\pi^2}{4}  \right) $   {\rule[-4.5ex]{0pt}{0pt}} \\ \hline

{\rule{0pt}{10ex}} 
Beta $BE(a, b)$  & $\begin{aligned}[c] 
& \frac{\Gamma(a + b)}{\Gamma(a) \Gamma( b)} x^{a - 1}(1-x)^{b -1} ,\\
& x \in (0,\: 1), a> 0, b>0 \\
& \zeta_k = a+ b + k
\end{aligned}$  & $ \frac{a}{\zeta_0} $   &$ \frac{a b}{\zeta_0^2\zeta_1} $   &  $ \frac{ 2a b(b - a) } {\zeta_0^3\zeta_1\zeta_2 } $ &  $\frac{ 3a b( 2(b - a)^2 + ab\zeta_2 } {\zeta_0^4\zeta_1 \zeta_2 \zeta_3 }  $    {\rule[-8ex]{0pt}{0pt}}  \\   \hline

{\rule{0pt}{7ex}} 
Binomial $B(n, p)$  & $\begin{aligned}[c] 
& \left(  \begin{matrix}
n \\ k
\end{matrix} \right) p^k (1-p)^{n-k},\\
& p \in [0,\: 1], k = 0, 1, 2, \cdots, n
\end{aligned}$  & $np$ & $\begin{aligned}
np(1-p)
\end{aligned} $  & $\begin{aligned}
& np(1-p)(1-2p)
\end{aligned} $  & $\begin{aligned}
& np(1-p)(1+ \\
& p(1-p)(3n-6))
\end{aligned} $    {\rule[-5.5ex]{0pt}{0pt}} \\  \hline

{\rule{0pt}{8ex}} 
Poisson $P(\lambda)$  & $\begin{aligned}[c] 
& \frac{\lambda^{k}}{k!} e^{-\lambda}, \lambda>0,\\
&     k = 0, 1, 2, \cdots,\infty
\end{aligned}$  & $\lambda $ & $\lambda $  & $\lambda $  & $\begin{aligned}
3\lambda^2 + \lambda   
\end{aligned} $    {\rule[-6ex]{0pt}{0pt}} \\   \hline

{\rule{0pt}{6ex}} 
Geometric $GE(p)$  & $\begin{aligned}[c] 
& p(1-p)^k , \:  p \in (0,\: 1], \\
&     k = 0, 1, 2, \cdots,\infty
\end{aligned}$  & $\begin{aligned}
\frac{(1-p)}{p} 
\end{aligned} $    & $\begin{aligned}
\frac{(1-p)}{p^2} 
\end{aligned} $    & $\begin{aligned}
\frac{(p-1)(p-2)}{p^3}   
\end{aligned} $   & $\begin{aligned}
\frac{(1-p)( p^2  - 9p  + 9)}{p^4}  
\end{aligned} $    {\rule[-4.2ex]{0pt}{0pt}} \\  \hline

{\rule{0pt}{10ex}}
$\begin{aligned}
&\mbox{Negative} \\
& \mbox{Binomial} \:NB(r, p)
\end{aligned} $ 
   & $\begin{aligned}[c] 
& \left(  \begin{matrix}
r + k -1 \\ k
\end{matrix} \right) p^r (1-p)^r,\\
&     k = 0, 1, 2, \cdots,\infty
\end{aligned}$  & $\begin{aligned}
\frac{r(1-p)}{p} 
\end{aligned} $    & $\begin{aligned}
\frac{r(1-p)}{p^2} 
\end{aligned} $    & $\begin{aligned}
\frac{r(p-1)(p-2)}{p^3}   
\end{aligned} $   & $ \frac{\begin{aligned}
 & r(1-p)( p^2  - 6p  \\
 &- 3pr  + 3r   + 6)  
\end{aligned} } {p^{-4}} $    {\rule[-4.5ex]{0pt}{0pt}} \\ 
\hline \hline 
\end{tabular}
\label{tab_probDistMoment}
\end{table*}

\IncMargin{1em}
\begin{algorithm}
\SetKwData{Left}{left}\SetKwData{This}{this}\SetKwData{Up}{up}
\SetKwFunction{Union}{Union}\SetKwFunction{FindCompress}{FindCompress}
\SetKwInOut{Input}{Note}\SetKwInOut{Output}{output}
\BlankLine
\emph{Prescribe the mean $\bar{\boldsymbol{x} }$, covariance $\boldsymbol{P}$, diagonal component of the skewness tensor $\breve{\boldsymbol{S}} =  [\begin{matrix} \boldsymbol{S}_{111}, \boldsymbol{S}_{222}, \cdots, \boldsymbol{S}_{nnn}  \end{matrix}  ]^T $, and the diagonal components of the kurtosis tensor $
    \breve{\boldsymbol{K}} =  [\begin{matrix} \boldsymbol{S}_{1111}, \boldsymbol{K}_{2222}, \cdots, \boldsymbol{K}_{nnnn}  \end{matrix}  ]^T$ }; 
\BlankLine
%
%
\emph{ Choose the free parameter vector $\boldsymbol{u} > \boldsymbol{0}$};
\BlankLine
\emph{Calculate the parameter vector $\boldsymbol{v}$ };
\begin{align*}
& \boldsymbol{v}  = \boldsymbol{u} +  \sqrt{\boldsymbol{P}}^{\odot -3}  \breve{\boldsymbol{S}}
\end{align*}
%
\BlankLine
\emph{Calculate the $2n+1$ sigma points};
\begin{align*}
\boldsymbol{\chi}_{[0]} & = \bar{\boldsymbol{x} } \qquad \qquad \qquad \quad   \boldsymbol{w}_0   \\
\boldsymbol{\chi}_{[i]} & = \bar{\boldsymbol{x} } - \boldsymbol{u}_i \sqrt{\boldsymbol{P} }_{[i]} \qquad  \:\:   \boldsymbol{w}'_i    \\
\boldsymbol{\chi}_{[i+n]} & = \bar{\boldsymbol{x} } + \boldsymbol{v}_{i}  \sqrt{\boldsymbol{P} }_{[i]}  \qquad \:\:   \boldsymbol{w}''_i   
\end{align*}
 for $i \in \{1, \cdots, n\}$, where $\sqrt{\boldsymbol{P} }_{[i]} $ is the $i$th column of the matrix square root of $\boldsymbol{P}$. 
 \BlankLine
\emph{Calculate the weights };
\begin{align*}
&  \boldsymbol{w}'' =  \boldsymbol{1} \oslash \boldsymbol{v} \oslash ( \boldsymbol{u} +   \boldsymbol{v} ) \\
& \boldsymbol{w}' =  \boldsymbol{w}'' \odot \boldsymbol{v} \oslash \boldsymbol{u},  \quad \boldsymbol{w}_0 = 1 - \sum_{i = 1}^{2n} \boldsymbol{w}_i
\end{align*}
where $\boldsymbol{w} = [\begin{matrix} \boldsymbol{w}_0, \boldsymbol{w}'^T, \boldsymbol{w}''^T  \end{matrix}]^T$
\BlankLine
\Input{To match the diagonal components of the kurtosis tensor, select {\tiny $     \boldsymbol{u} = \frac{1}{2} \left( -    \sqrt{\boldsymbol{P}}^{\odot -3} \breve{\boldsymbol{S}}   + \sqrt{ 4 \sqrt{\boldsymbol{P}}^{\odot -4}  \breve{\boldsymbol{K}}  - 3 \left(\sqrt{\boldsymbol{P}}^{\odot -3}  \breve{\boldsymbol{S}} \right)^{\odot 2} }  \right)   $ }in step 2.} 
\caption{Sigma Points for the Generalized Unscented Transform}\label{algo_disjdecomp}
\end{algorithm} \DecMargin{1em}

\section{Accuracy of Sigma Point Sample Statistics}
\label{section_sigmaAccuracy}
We demonstrate the accuracy of our sigma points in approximating any random vector $\boldsymbol{x}\in \mathbb{R}^n$. 
\begin{thm}
\label{Theorem_sampleStat}
Let $\boldsymbol{x}\in \mathbb{R}^n$ be any random vector with mean $\bar{\boldsymbol{x} }$ and covariance matrix $\boldsymbol{P}$, skewness tensor $\boldsymbol{S}$, and kurtosis tensor $\boldsymbol{K}$. The following statements are true for the $2n+1$ sigma points be defined as shown in Algorithm \ref{algo_disjdecomp}.
\begin{enumerate}
\item The sample mean, $\hat{\bar{\boldsymbol{x}}} = \sum_{i = 0}^{2n} \boldsymbol{w}_i \boldsymbol{\chi}_{[i]}$ is equal to $\bar{\boldsymbol{x}}$.
\item The sample covariance matrix, $\hat{\boldsymbol{P}} = \sum_{i = 0}^{2n} \boldsymbol{w}_i (\boldsymbol{\chi}_{[i]} - \hat{\bar{\boldsymbol{x}}})(\boldsymbol{\chi}_{[i]} - \hat{\bar{\boldsymbol{x}}})^T$, is equal to $\boldsymbol{P}$.
\item The sample skewness tensor $\hat{\boldsymbol{S}}_{jkl}= \sum_{i = 1}^{2n} \boldsymbol{w}_i (\boldsymbol{\chi}_{[i]} - \hat{\bar{\boldsymbol{x}}})_j (\boldsymbol{\chi}_{[i]} - \hat{\bar{\boldsymbol{x}}})_k (\boldsymbol{\chi}_{[i]} - \hat{\bar{\boldsymbol{x}}})_l  $, is equal to $\boldsymbol{S}_{jkl}$ if $j = k = l$.
\item The sample kurtosis tensor $\hat{\boldsymbol{K}}_{jklm} = \sum_{i = 1}^{2n} \boldsymbol{w}_i (\boldsymbol{\chi}_{[i]} - \hat{\bar{\boldsymbol{x}}})_j (\boldsymbol{\chi}_{[i]} - \hat{\bar{\boldsymbol{x}}})_k (\boldsymbol{\chi}_{[i]} - \hat{\bar{\boldsymbol{x}}})_l (\boldsymbol{\chi}_{[i]} - \hat{\bar{\boldsymbol{x}}})_m$, is equal to $\boldsymbol{K}_{jklm}$ if $j = k = l = m$ whenever $\boldsymbol{u} = \frac{  -    \sqrt{\boldsymbol{P}}^{\odot -3} \breve{\boldsymbol{S}}   + \sqrt{ 4 \sqrt{\boldsymbol{P}}^{\odot -4}  \breve{\boldsymbol{K}}  - 3 \sqrt{\boldsymbol{P}}^{\odot -6}  \breve{\boldsymbol{S}}^{\odot 2} }      }{2} $.
\end{enumerate}
\end{thm}
\begin{proof}
For our proof, we introduce diagonal matrices $\boldsymbol{U}, \boldsymbol{V} \in \mathbb{R}^{n \times n}$ such that $\boldsymbol{U} =  \mbox{diag}(\boldsymbol{u})$ and $\boldsymbol{V} =  \mbox{diag}(\boldsymbol{v})$. In matrix form, we evaluate the sample mean as
\begin{align}
    \hat{\bar{\boldsymbol{x}}} &=   [ \begin{matrix} \bar{\boldsymbol{x}}, &   \bar{\boldsymbol{x}}-\sqrt{\boldsymbol{P} } \boldsymbol{U},  & \bar{\boldsymbol{x}}+\sqrt{\boldsymbol{P} }  \boldsymbol{V} \end{matrix}  ] [ \begin{matrix} \boldsymbol{w}_0, &   \boldsymbol{w}'^T,   & \boldsymbol{w}''^T  \end{matrix}  ]^T \nonumber \\
& = \bar{\boldsymbol{x} } \sum_{i = 0}^{2n} \boldsymbol{w}_i + \sqrt{\boldsymbol{P} } (\boldsymbol{V} \boldsymbol{w}'' - \boldsymbol{U} \boldsymbol{w}')   \nonumber \\
& = \bar{\boldsymbol{x} }  + \sqrt{\boldsymbol{P} } (\boldsymbol{w}'' \odot \boldsymbol{v}  - \boldsymbol{w}' \odot \boldsymbol{u} )  =   \bar{\boldsymbol{x} }
\end{align}
because $\sum_{i = 0}^{2n}\boldsymbol{w}_i = 1$ and $\boldsymbol{w}'' \odot \boldsymbol{v}  = \boldsymbol{w}' \odot \boldsymbol{u}$. We see that the sample mean equals the actual mean. Evaluating the sample covariance matrix, we get
\begin{align}
\hat{\boldsymbol{P}} &=   \sqrt{\boldsymbol{P}} [ \begin{matrix}       \boldsymbol{U},  &   \boldsymbol{V} \end{matrix}  ] 
\left[ \begin{matrix}       \mbox{diag}(\boldsymbol{w}'),  &  \boldsymbol{0}   \\
\boldsymbol{0} , & \mbox{diag}(\boldsymbol{w}'') \end{matrix}  \right] 
\left[ \begin{matrix}      \boldsymbol{U} \sqrt{\boldsymbol{P} } \\    \boldsymbol{V} \sqrt{\boldsymbol{P} } \end{matrix}  \right]   \nonumber \\
&= \sqrt{\boldsymbol{P}}[ \begin{matrix}       \mbox{diag}(\boldsymbol{w}') \boldsymbol{U}^2 +    \mbox{diag}(\boldsymbol{w}'')\boldsymbol{V}^2 \end{matrix}  ]  \sqrt{\boldsymbol{P}}
\nonumber \\
& =  \sqrt{\boldsymbol{P}} \boldsymbol{I} \sqrt{\boldsymbol{P}} =   \boldsymbol{P} 
\end{align}
because $\boldsymbol{w}'\odot \boldsymbol{u}^{\odot 2} +   \boldsymbol{w}''\odot \boldsymbol{v}^{\odot 2} = \boldsymbol{1}$ is the diagonal of $\mbox{diag}(\boldsymbol{w}') \boldsymbol{U}^2 + \mbox{diag}(\boldsymbol{w}'')\boldsymbol{V}^2$. We see that the sample covariance matrix equals the actual covariance matrix. Defining $\hat{\breve{\boldsymbol{S}}} \in \mathbb{R}^n$ as a vector containing the diagonal components of the sample skewness tensor such that $$\hat{\breve{\boldsymbol{S}}}  =  [\begin{matrix} \hat{\boldsymbol{S}}_{111}, \hat{\boldsymbol{S}}_{222}, \cdots, \hat{\boldsymbol{S}}_{nnn}  \end{matrix}  ]^T $$ we can evaluate the diagonal components of the sample skewness tensor as
\begin{align}
\hat{\breve{\boldsymbol{S}}} & =  ([ \begin{matrix} -\sqrt{\boldsymbol{P} } \boldsymbol{U},  &  \sqrt{\boldsymbol{P} }  \boldsymbol{V} \end{matrix}  ] )^{\odot 3} [ \begin{matrix}  \boldsymbol{w}'^T,   & \boldsymbol{w}''^T  \end{matrix}  ]^T \\
& = [ \begin{matrix} -\sqrt{\boldsymbol{P} }^{\odot 3} \boldsymbol{U}^{\odot 3},  & \sqrt{\boldsymbol{P} }^{\odot 3} \boldsymbol{V}^{\odot 3} \end{matrix}  ]  [  \begin{matrix}  \boldsymbol{w}'^T,   & \boldsymbol{w}''^T  \end{matrix}  ]^T\\
& =  \sqrt{\boldsymbol{P} }^{\odot 3} [ - \begin{matrix} \boldsymbol{w}' \odot \boldsymbol{u}^{\odot 3}   + \boldsymbol{w}'' \odot \boldsymbol{v}^{\odot 3}   \end{matrix}  ]   \nonumber \\
& = \sqrt{\boldsymbol{P} }^{\odot 3} \sqrt{\boldsymbol{P}}^{\odot -3}  \breve{\boldsymbol{S}}   = \breve{\boldsymbol{S}}   \label{eqn_samp_skewness}
\end{align}
We see that our sigma points accurately match the diagonal components of the skewness tensor. Finally, defining $\hat{\breve{\boldsymbol{K}}} \in \mathbb{R}^n$ as a vector containing the diagonal components of the sample kurtosis tensor such that $$\hat{\breve{\boldsymbol{K}}}  =  [\begin{matrix} \hat{\boldsymbol{K}}_{1111}, \hat{\boldsymbol{K}}_{2222}, \cdots, \hat{\boldsymbol{K}}_{nnnn}  \end{matrix}  ]^T $$ we can evaluate the diagonal components of the sample kurtosis tensor as
\begin{align}
\hat{\breve{\boldsymbol{K}}} & =  ([ \begin{matrix} -\sqrt{\boldsymbol{P} } \boldsymbol{U},  &  \sqrt{\boldsymbol{P} }  \boldsymbol{V} \end{matrix}  ] )^{\odot 4} [ \begin{matrix}  \boldsymbol{w}'^T   & \boldsymbol{w}''^T  \end{matrix}  ]^T \\
& =  \sqrt{\boldsymbol{P} }^{\odot 4} [  \begin{matrix} \boldsymbol{w}' \odot \boldsymbol{u}^{\odot 4}   + \boldsymbol{w}'' \odot \boldsymbol{v}^{\odot 4}   \end{matrix}  ]   \nonumber \\
& = \sqrt{\boldsymbol{P} }^{\odot 4} \sqrt{\boldsymbol{P}}^{\odot -4}  \breve{\boldsymbol{K}}   = \breve{\boldsymbol{K}}   \label{eqn_samp_kurtosis}
\end{align}
We see that our sigma points accurately match the diagonal components of the kurtosis tensor.
\end{proof}
 Theorem~\ref{Theorem_sampleStat} shows that our sigma points in Algorithm~\ref{algo_disjdecomp} can accurately approximate the mean and covariance of any random vector, as well as the diagonal components of the skewness and kurtosis tensors -- this makes it applicable to a wide variety of applications.

\section{Constrained Sigma Points}
\label{section_sigmaConstrained}
Noting that several physical systems require some constraints on their states or parameters, we show how our sigma points can be constrained while at least maintaining second-order accuracy.

We require the sigma points to be constrained such that 
\begin{align*}
     \boldsymbol{a}  < \boldsymbol{\chi}_{[i]} < \boldsymbol{b}  \qquad  \mbox{for}  \:\: i \in \{0, \cdots, 2n \}  
\end{align*}
where $ \boldsymbol{a}  \in \mathbb{R}^n$ and $ \boldsymbol{b}\in \mathbb{R}^n$ are the lower bounds and upper bounds respectively.
\begin{asum}
The mean $\bar{\boldsymbol{x} }$ is within the bounds, such that $\boldsymbol{a} < \bar{\boldsymbol{x} } < \boldsymbol{b} $ 
\end{asum}

We note that our sigma points of Algorithm \ref{algo_disjdecomp} can violate some state constraints despite being able to accurately capture the mean and covariance of a random vector, as well as the diagonal components of its skewness and kurtosis tensors. This might make them inapplicable in situations/models that only permit constrained values. For example, in applications that assume a Poisson distribution for the states, such as count data, the states are usually positive by default and can never be negative. When our sigma point of Algorithm~\ref{algo_disjdecomp} is applied, the positive constraint on an independent random vector can be violated. We demonstrate this using the following example.
\begin{examp}
\label{example_alg1}
We generate sigma points for an independent Poisson random vector $x$ such that 
$$ \bar{\boldsymbol{x} } = \left[ \begin{matrix} 1.5\\ 1 \end{matrix}  \right],  
P  = \left[ \begin{matrix} 1.5, & 0\\ 0, & 1 \end{matrix}  \right],  
\breve{\boldsymbol{S}} = \left[ \begin{matrix} 1.5\\ 1 \end{matrix}  \right],  
\breve{\boldsymbol{K}} = \left[ \begin{matrix} 8.25\\ 4 \end{matrix}  \right]  $$
where $\bar{\boldsymbol{x} }$ is the mean, $P$ is the covariance matrix, and $\breve{\boldsymbol{S}}$ and $\breve{\boldsymbol{K}}$ are vectors containing the diagonal components of the skewness tensor and kurtosis tensor respectively. Using Algorithm~\ref{algo_disjdecomp}, we see that $\boldsymbol{w}_0 = 0.3333$, $\boldsymbol{w}_1 = 0.2049 $, $\boldsymbol{w}_2 = 0.2129 $, $\boldsymbol{w}_3 = 0.1284$, $\boldsymbol{w}_4 = 0.1204$, $\boldsymbol{u}_1 = 1.3713$, $\boldsymbol{u}_2 = 1.3028 $, $\boldsymbol{v}_1 = 2.1878$, and $\boldsymbol{v}_2 = 2.3028$. The $2n+1$ sigma points in matrix form is
\begin{align*}
 \boldsymbol{\chi} = \left[ \begin{matrix} 1.5,  &   -0.1794, &   1.5, &    4.1794, &   1.5  \\ 1,   & 1, &   -0.3028,  &  1, &  3.3028
 \end{matrix}  \right]   
\end{align*}
The sample statistics are $$ \hat{\bar{\boldsymbol{x}}}  = \left[ \begin{matrix} 1.5\\ 1 \end{matrix}  \right],   \hat{\boldsymbol{P}}  = \left[ \begin{matrix} 1.5, & 0\\ 0, & 1 \end{matrix}  \right], 
\hat{\breve{\boldsymbol{S}}} = \left[ \begin{matrix} 1.5\\ 1 \end{matrix}  \right],  
\hat{\breve{\boldsymbol{K}}}= \left[ \begin{matrix} 8.25\\ 4 \end{matrix}  \right] $$
\end{examp}
We see from Example~\ref{example_alg1} that despite the accuracy of the sample statistics, the sigma points $\boldsymbol{\chi}_{[1]}$ and $\boldsymbol{\chi}_{[2]}$  both had a negative value which do not satisfy the non-negativity of Poisson draws. 
%
\begin{cor}
\label{Theorem_lower}
  If the bound $\boldsymbol{a} < \bar{\boldsymbol{x} }$ is violated after implementing Algorithm \ref{algo_disjdecomp}, then enforcing the constraint $\boldsymbol{a} < \bar{\boldsymbol{x} }$  leads to accuracy in capturing only the mean, covariance matrix, and the diagonal components of the skewness tensor.
\end{cor}
\begin{proof}
Lower bounding $\boldsymbol{x}$ will require redefining the variable $\boldsymbol{u}$ such that (\ref{eqn_sConstrained}) is no longer satisfied. Theorem \ref{Theorem_sampleStat} establishes that violating  (\ref{eqn_sConstrained}) ensures an inaccurate approximation of the diagonal components of the kurtosis tensor. 
\end{proof}
\begin{cor}
\label{Theorem_upper}
  If the bound $  \bar{\boldsymbol{x} } < \boldsymbol{b} $ is violated after implementing Algorithm \ref{algo_disjdecomp},  then enforcing either $ \bar{\boldsymbol{x} } < \boldsymbol{b} $ or $  \boldsymbol{a} < \bar{\boldsymbol{x} } < \boldsymbol{b}$ leads to accuracy in capturing only the mean and covariance matrix.
\end{cor}
\begin{proof}
Both cases,  $ \bar{\boldsymbol{x} } < \boldsymbol{b} $ or $  \boldsymbol{a} < \bar{\boldsymbol{x} } < \boldsymbol{b}$, require the redefinition of the variable $\boldsymbol{v}$. This means that (\ref{eqn_ndim_ssi}) will no longer be satisfied. Theorem \ref{Theorem_sampleStat} establishes that violating  (\ref{eqn_ndim_ssi}) ensures an inaccurate approximation of the diagonal components of skewness and kurtosis tensor. 
\end{proof}

To enforce constraints on the sigma points, we introduce a \textit{slack parameter} $\theta \in (0, \cdots, 1)$ which is a user selected constant. Using $\theta$, we now redefine the free parameters $\boldsymbol{u}_i$ and $\boldsymbol{v}_i$ for $i \in  \{1, \cdots, n\}$ as
\begin{align*}
    \boldsymbol{u}_i &= \theta  \left[\min \left\{ \left| (\bar{\boldsymbol{x} } - \boldsymbol{a}) \oslash  \sqrt{P }_{[i]}  \right|  \right\} \right]\quad \mbox{if}  \:   \boldsymbol{\chi}_{[i]}  <  \boldsymbol{a} \\
    \boldsymbol{v}_i &= \theta  \left[\min \left\{ \left| (\boldsymbol{b} - \bar{\boldsymbol{x} } ) \oslash \sqrt{P }_{[i]}  \right|  \right\} \right]\quad \mbox{if}  \:   \boldsymbol{\chi}_{[i+n]}  > \boldsymbol{b}
\end{align*}
where $|.|$ denotes the absolute value, and the sigma points get closer to their constraints as $\theta \rightarrow 1$. We note that the equations for $\boldsymbol{w}'$ and $\boldsymbol{w}''$ are unchanged. 

We note that enforcing constrains on the sigma points results in a loss of accuracy in capturing the diagonal components of at least the kurtosis tensor. The constrained sigma point algorithm is given in Algorithm~\ref{algo_disjdecomp2}. We now show a benefit of Algorithm~\ref{algo_disjdecomp2} in the following example.

\IncMargin{1em}
\begin{algorithm}
\SetKwData{Left}{left}\SetKwData{This}{this}\SetKwData{Up}{up}
\SetKwFunction{Union}{Union}\SetKwFunction{FindCompress}{FindCompress}
\SetKwInOut{Input}{Note}\SetKwInOut{Output}{output}
\BlankLine
\emph{Implement Algorithm~\ref{algo_disjdecomp}  }
\BlankLine
\If{$  \boldsymbol{\chi}_{[i]}   < \boldsymbol{a}$ for $i\in \{1, \cdots,  2n\}$ }{
\BlankLine
\If{$    i  \leq n $ }{
$\boldsymbol{u}_i = \theta \left[\min \left\{ \left| (\bar{\boldsymbol{x}} - \boldsymbol{a} ) \oslash \sqrt{P }_{[i]}  \right|  \right\} \right] $
}
\If{$    i  > n $ }{
$\boldsymbol{v}_{i-n} = \theta \left[\min \left\{ \left| (\boldsymbol{a} - \bar{\boldsymbol{x}} ) \oslash \sqrt{P }_{[i-n]}  \right|  \right\} \right] $
}
}
\BlankLine
\emph{Repeat steps 3 and 4 of Algorithm~\ref{algo_disjdecomp} if $\boldsymbol{v}$ was not redefined, otherwise repeat only step 4 of Algorithm~\ref{algo_disjdecomp}};
\BlankLine
\If{$  \boldsymbol{\chi}_{[i]}  > \boldsymbol{b}$ \mbox{for} $i\in \{1, \cdots,  2n\}$ }{
\BlankLine
\If{$    i  \leq n $ }{
$\boldsymbol{u}_i = \theta \left[\min \left\{ \left| (\bar{\boldsymbol{x}} - \boldsymbol{b} ) \oslash \sqrt{P }_{[i]}  \right|  \right\} \right] $
}
\If{$    i  > n $ }{
$\boldsymbol{v}_{i-n} = \theta \left[\min \left\{ \left| (\boldsymbol{b} - \bar{\boldsymbol{x}} ) \oslash \sqrt{P }_{[i-n]}  \right|  \right\} \right] $
}
}

\BlankLine
\BlankLine
\emph{Repeat steps 3, 4, and 5 of Algorithm~\ref{algo_disjdecomp} if $\boldsymbol{v}$ was not redefined, otherwise repeat only steps 4 and 5 of Algorithm~\ref{algo_disjdecomp}};
\BlankLine
\Input{$\theta \in (0, \cdots, 1)$ is a user defined constant. The sigma points get closer to their constraints as $\theta \rightarrow 1$.} 
\caption{Constrained Sigma Points for the Generalized Unscented Transform}\label{algo_disjdecomp2}
\end{algorithm} \DecMargin{1em}

\begin{figure}
\centering     
\subfigure[]{\label{fig_sigma}\includegraphics[width=70mm, height = 55mm]{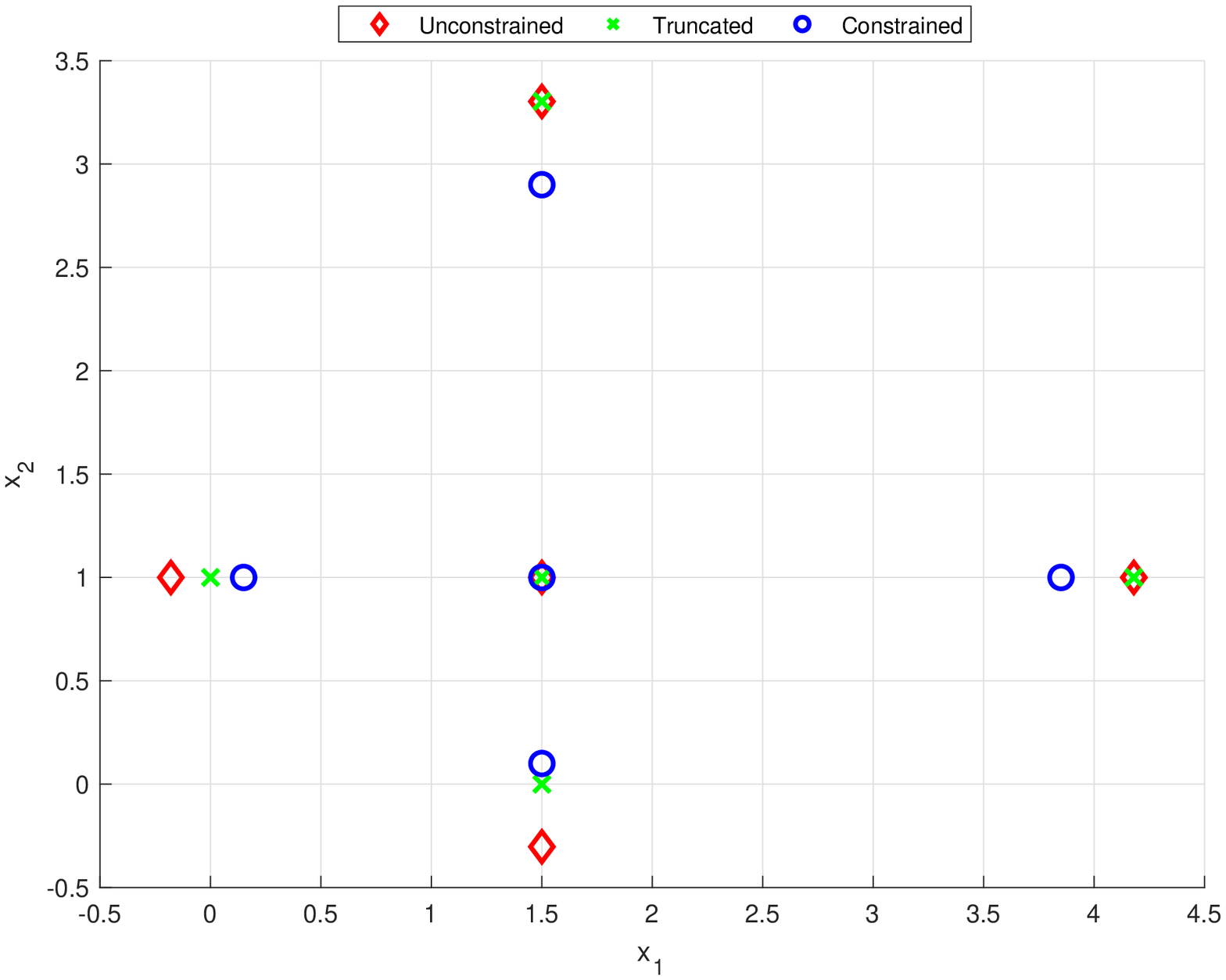}} \qquad
\subfigure[]{\label{fig_sigmaCov}\includegraphics[width=80mm, height = 55mm]{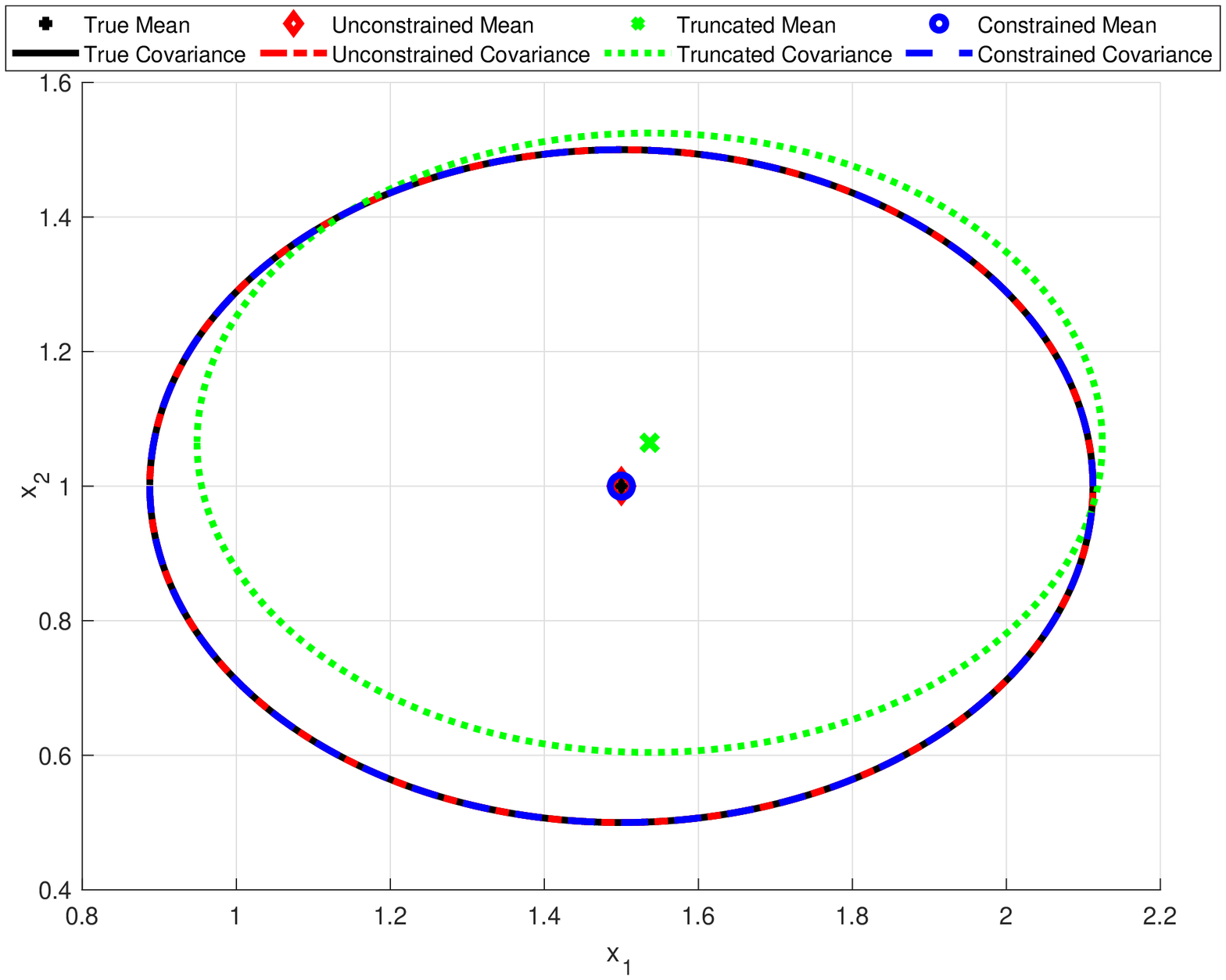}}
\caption{(a) Locations of sigma points for the unconstrained (Algorithm~\ref{algo_disjdecomp}), truncated, and constrained (Algorithm~\ref{algo_disjdecomp2}) sigma points. (b) Mean and covariance of the unconstrained (Algorithm~\ref{algo_disjdecomp}), truncated, and constrained (Algorithm~\ref{algo_disjdecomp2}) sigma points.}
\label{fig_Example4}
\end{figure}

\begin{examp}
\label{example_alg2}
Using Algorithm~\ref{algo_disjdecomp2} to generate positively constrained sigma points for the Poisson random vector, we select $\theta = 0.9$. We see that $\boldsymbol{w}_0 = -0.0576$, $\boldsymbol{w}_1 = 0.3003$, $\boldsymbol{w}_2 = 0.3968$, $\boldsymbol{w}_3 = 0.1725$, $\boldsymbol{w}_4 = 0.188$, $\boldsymbol{u}_1 = 1.1023$, $\boldsymbol{u}_2 = 0.9$, $\boldsymbol{v}_1 = 1.9188$, and $\boldsymbol{v}_2 = 1.9$. The $2n+1$ positive sigma points in matrix form is
\begin{align*}
 \boldsymbol{\chi} = \left[ \begin{matrix} 1.5,    & 0.15, &   1.5, &    3.85  &   1.5  \\1,  &  1, &   0.1, &   1, &  2.9 
 \end{matrix}  \right]    
\end{align*}
while the corresponding sample statistics are $$ \hat{\bar{\boldsymbol{x}}}  = \left[ \begin{matrix} 1.5\\ 1 \end{matrix}  \right],   \hat{\boldsymbol{P}}  = \left[ \begin{matrix} 1.5, & 0\\ 0, & 1 \end{matrix}  \right], 
\hat{\breve{\boldsymbol{S}}} = \left[ \begin{matrix} 1.5\\ 1 \end{matrix}  \right],  
\hat{\breve{\boldsymbol{K}}}= \left[ \begin{matrix} 6.2587\\ 2.7100 \end{matrix}  \right]$$
\end{examp}
We see from Example~\ref{example_alg2} that using Algorithm~\ref{algo_disjdecomp2} ensures that the sigma points are always positive while ensuring accuracy in approximating the true mean and covariance of a random vector, as well as capturing the diagonal components of the skewness tensor. However, the ability to exactly capture the diagonal components of the kurtosis tensor is lost. A graphical representation of Examples~\ref{example_alg1} and ~\ref{example_alg2} is shown in Fig.~\ref{fig_Example4} where we plot the sigma points and the covariance.

\section{Propagation of Means and Covariances of Nonlinear Transformations}
\label{section_propagation}
We analyze the performance of our new sigma point algorithm when they undergo nonlinear transformations. We will show how linearization approximations, via Taylor series expansion of a nonlinear transformation of a random vector $\boldsymbol{x}$ evaluated about its mean $\bar{\boldsymbol{x} }$, introduce errors in the propagation of means and covariances. In Appendix \ref{appendix_true}, we evaluated the true mean and true covariance of a random vector, as well as the approximated mean and approximated covariance. We see that although errors are introduced beyond the third order when approximating a nonlinear transformation of a random vector, these errors are minimized because of our ability to match the diagonal components of the skewness and kurtosis tensors. We also see that errors are introduced beyond the third order when the random vector is independent.

We will see that errors can be introduced in the propagation of means and covariances beyond the second order when sigma points developed under the Gaussian assumption~\cite{julier1996general,julier1997consistent, julier2004unscented} are used to approximate the nonlinear function $\boldsymbol{\lambda}(\boldsymbol{x})$ when $\boldsymbol{x}$ is an independent random vector. We note that the nonlinear transformation $\boldsymbol{y} \in \mathbb{R}^n$ is given by 
\begin{align}
\boldsymbol{y}  =  \boldsymbol{\lambda}(\boldsymbol{x}) \label{eqn_Ydyn}
\end{align}
where $\mathbb{E}[\boldsymbol{x}]  = \bar{\boldsymbol{x} }$. $\bar{\boldsymbol{x} }$, $\boldsymbol{P}$, $\breve{\boldsymbol{S}}$, and $\breve{\boldsymbol{K}}$, we evaluate the sample mean and covariance of the nonlinear transformation of~(\ref{eqn_Ydyn}) using Algorithm~\ref{algo_disjdecomp}.
    
For our comparison, we use the scaled unscented transform of~\cite{julier2004unscented}, which is denoted as UT for the remainder of this paper, and the higher order sigma point unscented transform (HOSPUT) of \cite{ponomareva2010new}. The scaling of the UT was selected to match a Gaussian distribution. We do not compare against the sigma points in \cite{straka2012randomized, rezaie2016skewed, hou2019high, easley2021higher} because they either use a Gaussian assumption, a closed skew normal distribution, or more than $2n+1$ sigma points. The sample mean and sample covariance can be evaluated using (\ref{eqn_tr_step2})--(\ref{eqn_tr_step4}).

\subsection{Case Study 1 -- Transformation of Random Variables}
\label{section_caseStudy}
Defining $x$ as a random variable that can follow any of the probability distributions given in the Table \ref{tab_probDistMoment}, we evaluate the sample mean and covariance of two nonlinear transformations: a quadratic function of the random variable $y = 3x + 2x^2$, and a trigonometric function of the random variable $y = \sin(x)$. We also use $10^5$ Monte Carlo draws from the different probability distributions. The true mean and covariance of the quadratic function can be easily evaluated using the raw moments of $x$ up to its fourth order. The true mean and covariance of the trigonometric function can be evaluated using their characteristic functions. A comparison between the accuracy of the GenUT, UT, $10^5$ Monte Carlo draws, and HOSPUT in approximating the true mean and true covariance of the nonlinear transformations for the different probability distributions is shown in Tables~\ref{tab_probExample1Mean}-\ref{tab_probExample2Cov}. 

For the quadratic function, we see that both the GenUT and HOSPUT gave an exact approximation of the true mean and true covariance for all the probability distributions while the UT was only accurate in approximating the true mean when the probability distribution was not Gaussian. This is because the GenUT and HOSPUT are accurate up to the fourth order moments when the random variable $x$ has a dimension of 1. Although the $10^5$ Monte Carlo draws gave relatively good approximations, they were not as accurate as the GenUT.

\begin{figure}
\centering     
\subfigure[]{\label{fig:poiss}\includegraphics[width = 0.32\textwidth,height = 5cm]{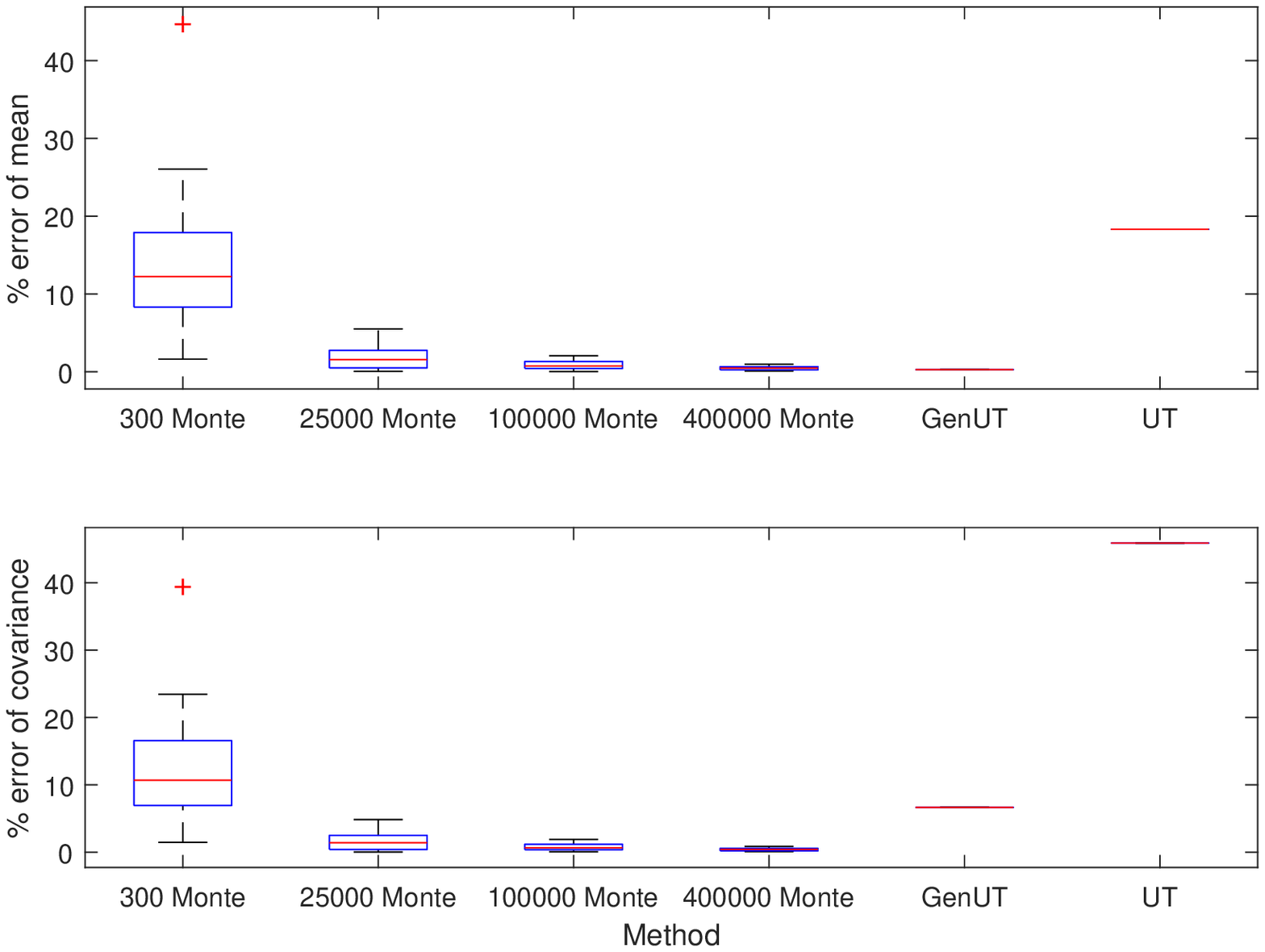}}  \hfill
\subfigure[]{\label{fig:weibull}\includegraphics[width = 0.32\textwidth,height = 5cm]{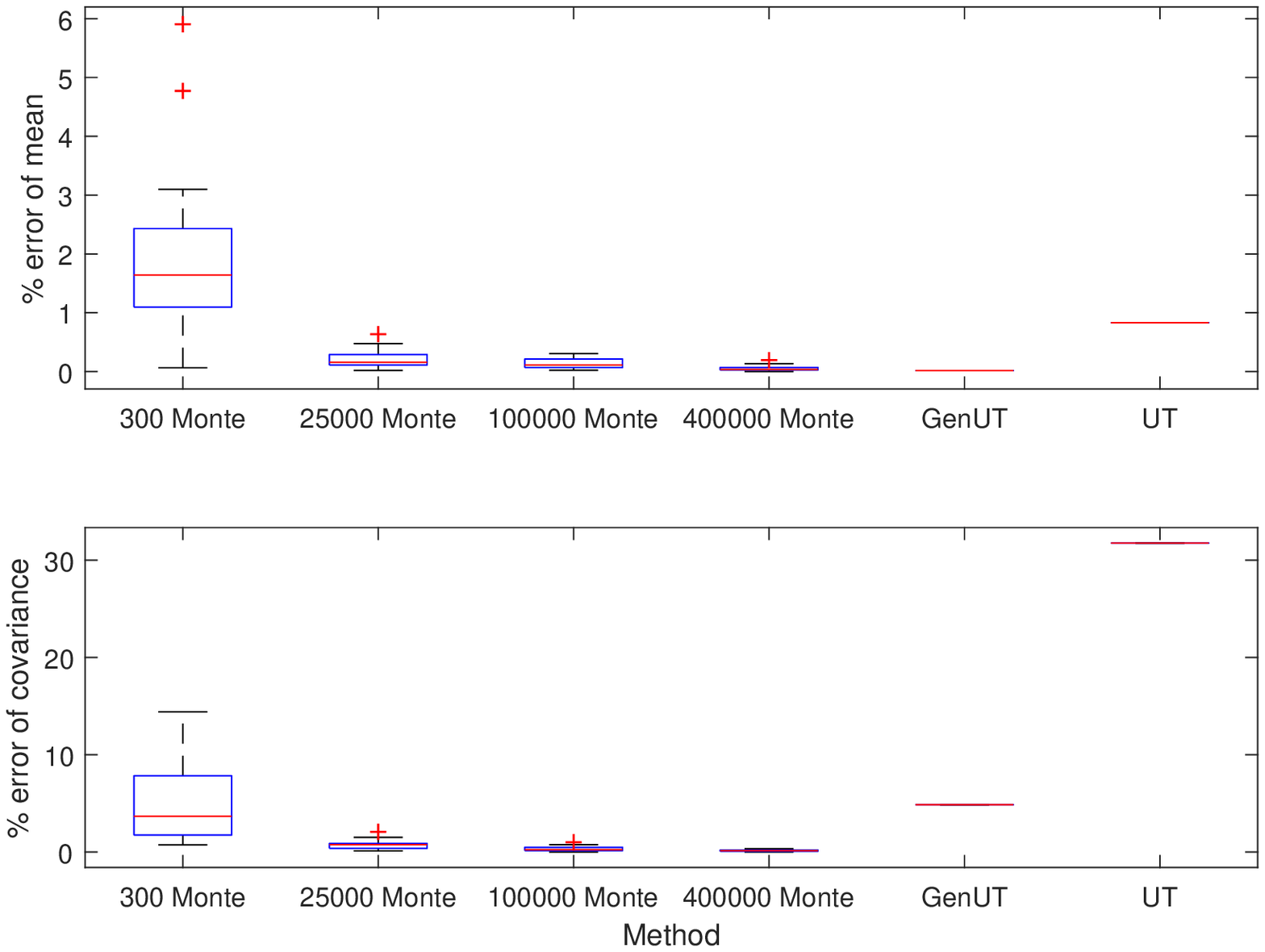}} \hfill  
\caption{(a) Moments of $y = \sin(x)$ when $x$ is a Poisson random variable. (b) Moments of $y = \sin(x)$ when $x$ is a Weibull random variable.}
\label{fig_boxPlot}
\end{figure}

For the trigonometric function, we see that the GenUT, HOSPUT, and UT were unable to give exact approximations of the true mean and true covariance in most cases because the Taylor series expansion of $\boldsymbol{\lambda}(\boldsymbol{x})$  has terms beyond the fourth order. The GenUT and HOSPUT were more accurate than the UT for all the non-Gaussian probability distributions because they are both accurate up to the fourth order while the UT is accurate up to the second order. The $10^5$ Monte Carlo draws sometimes gave better accuracy than the GenUT because of the random nature of its draws. A box plot of the accuracy of the GenUT, UT, and several Monte Carlo draws of different sizes is shown in Fig. \ref{fig_boxPlot} for the trigonometric function. We do not include the HOSPUT because it gives the same performance as the GenUT when a single random variable is transformed. We see that a significant number of Monte Carlo draws is needed to achieve the accuracy of the GenUT when approximating the mean. A significant number of Monte Carlo draws gives better accuracy in approximating the variance. 

 \begin{table} 
\small
\caption{Percentage error in Propagating the mean of $y = 3x + 2x^2$ } 
\centering 
\begin{tabular}{ c c c c c} 
\hline\hline 
 $x$ & GenUT   & UT  &  MC & HOSPUT    \\
\hline \hline 
$\mathcal{N}(1, 4)$    &        0    &     0 &   0.015 & 0 {\rule[2.5ex]{0pt}{0pt}} \\ \hline
$E(2)$         &    0   &      0 &   0.069  & 0
{\rule[2.5ex]{0pt}{0pt}}  \\ \hline
$G(1,2)$  &    0 &        0  &   0.452  & 0
{\rule[2.5ex]{0pt}{0pt}}  \\ \hline
 $W(1, 2)$     &     0 &   0  &  0.005 & 0
   {\rule[2.5ex]{0pt}{0pt}} \\ \hline
$R(1)$   &    0   &      0  &  0.097   & 0
{\rule[2.5ex]{0pt}{0pt}}  \\ \hline 
$BE(3,4)$    &      0    &     0  &  0.063  & 0 
{\rule[2.5ex]{0pt}{0pt}}  \\ \hline
$B(3,0.3)$    &    0   & 0   &   0.457  & 0
{\rule[2.5ex]{0pt}{0pt}}  \\ \hline
$P(2)$  &        0   &      0  &  0.270 & 0
  {\rule[2.5ex]{0pt}{0pt}}  \\ \hline 
$GE(0.5)$   &         0   &      0  &  1.251  & 0
{\rule[2.5ex]{0pt}{0pt}}  \\ \hline
$NB(4, 0.67)$    & 0   & 0  &    0.668   & 0
{\rule[2.5ex]{0pt}{0pt}}  \\ 
\hline \hline 
\end{tabular}
\label{tab_probExample1Mean}
\end{table}

 \begin{table} 
\small
\caption{Percentage error in Propagating the covariance of $y = 3x + 2x^2$ } 
\centering 
\begin{tabular}{ c c c c  c} 
\hline\hline 
 $x$ & GenUT   & UT  &  MC  & HOSPUT    \\
\hline \hline 
$\mathcal{N}(1, 4)$        &       0    &     0  &  0.029  & 0 {\rule[2.5ex]{0pt}{0pt}} \\ \hline
$E(2)$     & 0 &  49.057  &  0.249  & 0
{\rule[2.5ex]{0pt}{0pt}} \\ \hline
$G(1,2)$  & 0 &  64 &   1.889   & 0
{\rule[2.5ex]{0pt}{0pt}} \\ \hline
 $W(1, 2)$    &  0 &   15.003  &   0.310   & 0
 {\rule[2.5ex]{0pt}{0pt}} \\ \hline
$R(1)$   &  0  &  16.815 &   0.381  & 0
{\rule[2.5ex]{0pt}{0pt}} \\ \hline 
$BE(3,4)$     &      0  &  2.307   &  0.613  & 0
 {\rule[2.5ex]{0pt}{0pt}} \\ \hline
$B(3,0.3)$    &    0 &  16.380  &  0.359   & 0
{\rule[2.5ex]{0pt}{0pt}} \\ \hline
$P(2)$  &  0  &  25.946  &  1.061   & 0
 {\rule[2.5ex]{0pt}{0pt}} \\ \hline 
$GE(0.5)$    &   0 &  67.662  &  1.036  & 0
{\rule[2.5ex]{0pt}{0pt}} \\ \hline
$NB(4, 0.67)$    &   0  &   43.224  &    2.356   & 0 {\rule[2.5ex]{0pt}{0pt}}  \\
\hline \hline 
\end{tabular}
\label{tab_probExample1Cov}
\end{table}

\begin{table} 
\small
\caption{Percentage error in Propagating the mean of $y = \sin(x)$ } 
\centering 
\begin{tabular}{ c c c c c} 
\hline\hline 
 $x$ & GenUT   & UT  &  MC  & HOSPUT   \\
\hline \hline 

 $\mathcal{N}(1.57, 0.1)$     &    0.001 &    0.001 &    0.012  &    0.001   {\rule[2.5ex]{0pt}{0pt}}  \\ \hline
$E(2)$     &    0.219 &    5.788 &    0.110   &    0.219  {\rule[2.5ex]{0pt}{0pt}}  \\ \hline
$G(0.5,0.5)$    &    0.312 &    6.964 &    0.050  &    0.312
   {\rule[2.5ex]{0pt}{0pt}}  \\ \hline
$W(1,2)$    &   0.017 &   0.831 & 0.029    &   0.017
{\rule[2.5ex]{0pt}{0pt}} \\ \hline
$R(1)$    &    0.049 &    0.912 & 0.007   &    0.049 
{\rule[2.5ex]{0pt}{0pt}} \\ \hline
$BE(3,4)$    &    0  &   0.038 &    0.037 &    0
   {\rule[2.5ex]{0pt}{0pt}}  \\ \hline
$B(3,0.3)$    &    0.158 &  4.814 & 0.046   &    0.158
{\rule[2.5ex]{0pt}{0pt}}  \\ \hline
$P(0.1)$    &    0.275 &   18.305 & 0.531  &    0.275
{\rule[2.5ex]{0pt}{0pt}}  \\ \hline
$GE(0.7)$     &    2.416  &   32.906 & 0.138    &    2.416 {\rule[2.5ex]{0pt}{0pt}}  \\ \hline
$NB(0.4,0.67)$    &    0.176 &   44.172 & 0.383   &    0.176    {\rule[2.5ex]{0pt}{0pt}}  
\\ \hline \hline 
\end{tabular}
\label{tab_probExample2Mean}
\end{table}

\begin{table} 
\small
\caption{Percentage error in Propagating the covariance of $y = \sin(x)$ } 
\centering 
\begin{tabular}{ c c c c c} 
\hline\hline 
 $x$ & GenUT   & UT  &  MC  & HOSPUT   \\
\hline \hline 

 $\mathcal{N}(1.57, 0.1)$     & 5.026    & 5.026  &   0.444  & 5.026 {\rule[2.5ex]{0pt}{0pt}} \\ \hline
$E(2)$      & 23.499 &   72.557 &    0.213   & 23.499
{\rule[2.5ex]{0pt}{0pt}} \\ \hline
$G(0.5,0.5)$     &   20.749 &   61.391 &    0.372  &   20.749  {\rule[2.5ex]{0pt}{0pt}} \\ \hline
$W(1,2)$    & 4.862 &  31.760 &   0.043 & 4.862
{\rule[2.5ex]{0pt}{0pt}} \\ \hline
$R(1)$     & 12.158 &   50.678 &   0.531   & 12.158
{\rule[2.5ex]{0pt}{0pt}} \\ \hline
$BE(3,4)$     & 0.031 &   0.940  &    0.225  & 0.031
{\rule[2.5ex]{0pt}{0pt}} \\ \hline
$B(3,0.3)$    & 11.033 &   24.806 &    0.060  & 11.033
{\rule[2.5ex]{0pt}{0pt}} \\ \hline
$P(0.1)$     & 6.646  &   45.895 &    0.461    & 6.646
{\rule[2.5ex]{0pt}{0pt}} \\ \hline
$GE(0.7)$      & 12.074 &  87.637 &    0.070   & 12.074
{\rule[2.5ex]{0pt}{0pt}} \\ \hline
$NB(0.4,0.67)$     & 39.068 &  135.783  &   0.366  & 39.068 {\rule[2.5ex]{0pt}{0pt}}
\\ \hline \hline 
\end{tabular}
\label{tab_probExample2Cov}
\end{table}

\subsection{Case Study 2 -- Transformation of a Random Vector }
\label{section_caseStudy-2}
We examine the performance of the GenUT, HOSPUT, and UT in approximating the true mean and covariance of a nonlinear transformation of different random variables such that
\begin{align}
\boldsymbol{x} = \left[\begin{matrix}
x_1\\ x_2
\end{matrix}   \right] =  \left[\begin{matrix}
\mbox{Poisson}(0.1) \\ \mbox{Rayleigh}(1)
\end{matrix}   \right], \quad 
\boldsymbol{Y} = \left[\begin{matrix}
\sin(\boldsymbol{x}_1 \boldsymbol{x}_2) \\ \cos(\boldsymbol{x}_1 \boldsymbol{x}_2)
\end{matrix}   \right]
\end{align}
We calculate the true mean and true covariance of $\boldsymbol{y}$ using $10^7$ Monte Carlo draws. The percentage error in approximating each element of the mean is
$$ \begin{matrix}
 \mbox{Mean} \\ \% \mbox{ error}
\end{matrix}   =   \underbrace{ \left[ \begin{matrix}
 24.7 \\ 0.05
\end{matrix}   \right] }_\text{GenUT}, \quad 
 \underbrace{ \left[ \begin{matrix}
63.87 \\ 1.43
\end{matrix}   \right] }_\text{UT}, \quad   
\underbrace{ \left[ \begin{matrix}
51.64 \\ 1.23
\end{matrix}   \right] }_\text{HOSPUT}  $$
The percentage error in approximating each element of the covariance matrix is
 \begin{align*}
 \begin{matrix}
 \mbox{Covariance} \\ \% \mbox{ error}
\end{matrix}   =    &\underbrace{ \left[ \begin{matrix}
24.68, &   8.77  \\ 8.77,  &  20.13
\end{matrix}   \right] }_\text{GenUT }, \quad 
 \underbrace{ \left[ \begin{matrix}
145.51, &  68.47  \\ 68.47, &   83.16 
\end{matrix}   \right] }_\text{UT}\\   
& \underbrace{ \left[ \begin{matrix}
126.93, &  28.51  \\ 28.51, &  71.72 
\end{matrix}   \right] }_\text{HOSPUT}  
\end{align*}
We see that for the nonlinear transformation, the GenUT gave the lowest percentage error when approximating the elements of the mean and covariance matrix. The UT gave the worst performance because it was unable to account for the non-Gaussian distributed nature of the random variable $\boldsymbol{x}$. The HOSPUT performed worse than the GenUT because, when the problem dimension exceeds 1, it is only able to match the average values of the diagonal elements of the skewness and kurtosis tensors.

\subsection{Case Study 3 - Infectious Disease Models}
\label{section_caseStudy_3} 
We consider an SIR (susceptible-infectious-recovered) infectious disease model given by the difference equation \cite{keeling2011modeling}
\begin{align}
S_{k+ 1} & = S_k -  \frac{\beta S_k I_k}{N}     \nonumber \\
I_{k+ 1} & = I_t +   \frac{\beta S_k I_k}{N}   -   \gamma I_k     \label{eqn_SIR_det} \\
R_{k+ 1} & = R_k +  \gamma I_k     \nonumber     
\end{align}
where $\beta$ is the infection rate, $\gamma$ is the recovery rate, and $N = S_k + I_k + R_k$. Using the conservation principle $S + I + R = N$, we reduce the model of (\ref{eqn_SIR_det}) to
\begin{align*}
I_{k+ 1} & = I_t +   \beta\left(N - I_k - R_k  \right)\frac{I_k}{N} - \gamma I_k \\ 
R_{k+ 1} & = R_k +  \gamma I_k \nonumber     
\end{align*}
We note that by defining $\boldsymbol{x}   =  \mbox{Poisson} \left[\begin{matrix}
I_k  &
 R_k
\end{matrix}  \right]^T $, we can rewrite the above equation as
\begin{align}
\left[\begin{matrix}
I_{k+1} \\ R_{k+1}
\end{matrix}  \right] = \left[\begin{matrix}
I_k \\ R_k
\end{matrix}  \right] + \left[\begin{matrix}
\beta\left(N - \boldsymbol{x}_1 - \boldsymbol{x}_2  \right)\frac{\boldsymbol{x}_1}{N} \\ \gamma \boldsymbol{x}_1
\end{matrix}  \right] \label{eqn_SIR_poissReduced1}         
\end{align}
where $\boldsymbol{x}_{i}$ is the $i$th element of the vector $\boldsymbol{x}$. 

We examine the performance of the GenUT, HOSPUT, and UT in approximating the true mean and covariance of (\ref{eqn_SIR_poissReduced1}). We use the parameters $I_k = 10$, $R_k = 2$, $\beta = 1.5$, $\gamma = 0.3$, and $N = 100$.
The percentage error in approximating each element of the mean is
$$ \begin{matrix}
 \mbox{Mean} \\ \% \mbox{ error}
\end{matrix}   =   \underbrace{ \left[ \begin{matrix}
 0 \\ 0
\end{matrix}   \right] }_\text{GenUT}, \quad 
 \underbrace{ \left[ \begin{matrix}
0 \\ 0
\end{matrix}   \right] }_\text{UT}, \quad   
\underbrace{ \left[ \begin{matrix}
0 \\ 0
\end{matrix}   \right] }_\text{HOSPUT}  $$
The percentage error in approximating each element of the covariance matrix is
$$ \begin{matrix}
 \mbox{Covariance} \\ \% \mbox{ error}
\end{matrix}   =    \underbrace{ \left[ \begin{matrix}
0.03, & 0  \\ 0, & 0
\end{matrix}   \right] }_\text{GenUT }, \quad 
 \underbrace{ \left[ \begin{matrix}
2.56, &   1.3  \\ 1.3,   &  0
\end{matrix}   \right] }_\text{UT}, \quad   
\underbrace{ \left[ \begin{matrix}
0.3, &   0.13   \\ 0.13, & 0
\end{matrix}   \right] }_\text{HOSPUT}  $$
We see that the GenUT gave the least approximation error of the true covariance matrix. The inability of the GenUT to exactly match the true covariance matrix is because the GenUT is only able to capture the diagonal components of the skewness and kurtosis tensors.

\section{Conclusion}
\label{section_conclusion}
In this paper we have developed the generalized unscented transform (GenUT) that is capable of adapting to the unique statistics of an arbitrarily distributed random variable. We showed that due to its ability to match the diagonal elements of the skewness and kurtosis tensors of most random vectors using $2n+1$ sigma points, the GenUT is preferable to and more accurate than unscented transforms that were either developed using the Gaussian assumption or were developed without any probability distribution in mind. 

In terms of ease of implementation, we demonstrated that like the unscented transform originally developed in~\cite{julier1997consistent} which uses $2n + 1$ sigma points, the GenUT uses the same number of sigma points. When compared against unscented transforms that employ more than $2n+1$ sigma points, the GenUT is characterized by a lower computational cost due to its lower number of sigma points that scales linearly with the problem dimension.

In terms of performance,  the GenUT and unscented transforms that use $2n+1$ sigma points developed under the Gaussian assumption give the same performance when the random variable is Gaussian distributed. However, when the random variable or random vector is not Gaussian distributed, the GenUT gives better accuracy in the propagation of means and covariances. Additionally, we also showed that the GenUT formulation makes it easy to analytically enforce constraints on the sigma points while still guaranteeing at least a second-order accuracy, which makes it appealing in models that permit only constrained values for random variables or parameters. 

For uncertainty quantification, estimation, or prediction applications, when compared to existing unscented transforms, the GenUT gives the most accuracy that can be gotten by employing $2n +1$ sigma points. This accuracy will have more significant consequences if the nonlinearities are strong and the problem dimension is large. The GenUT can be applied to any filter that uses linear or nonlinear transformations of random variables. The MATLAB\textsuperscript{\textregistered} source code used to
generate the results in this paper
is available at \cite{sourceGenUT}.

\appendices

\section{True Mean and Covariance of Nonlinear Transformations}
\label{appendix_true}
We derive analytical expressions for the true mean and covariance when we take the Taylor series expansion of the nonlinear function $y = \boldsymbol{\lambda}(\boldsymbol{x})$ where $\boldsymbol{x}$ is a random vector.  

\subsection{True Mean of the Nonlinear Transformation}
\label{Appendix_trueMean}
Applying Taylor series expansion around $\bar{\boldsymbol{x}}$, where $\tilde{\boldsymbol{x}} = \boldsymbol{x} - \bar{\boldsymbol{x}}$, we write the true mean of $\boldsymbol{y}$ as
\begin{align} 
\bar{\boldsymbol{y}} & =  \mathbb{E} \left[\boldsymbol{\lambda}(\boldsymbol{x}) \right] \nonumber \\
 & = \boldsymbol{\lambda}(\bar{\boldsymbol{x}}) + \mathbb{E}\left[ D_{\tilde{\boldsymbol{x}}} \boldsymbol{\lambda} + \frac{D_{\tilde{\boldsymbol{x}}}^2 \boldsymbol{\lambda}}{2!} + \frac{D_{\tilde{\boldsymbol{x}}}^3 \boldsymbol{\lambda}}{3!} + \frac{D_{\tilde{\boldsymbol{x}}}^4 \boldsymbol{\lambda}}{4!}+ \cdots \right]
 \label{eqn_ExpectedYtrue}
\end{align} 
where $D_{\tilde{\boldsymbol{x}}} \boldsymbol{\lambda}$ is the total differential of $\boldsymbol{\lambda}(\boldsymbol{x})$ when perturbed around a nominal value $\bar{\boldsymbol{x}}$ by $\tilde{\boldsymbol{x}}$. We note that 
\begin{align}
D_{\tilde{\boldsymbol{x}}}^k \boldsymbol{\lambda}  & = \left. \left( \sum_{i = 1}^{n}   \tilde{\boldsymbol{x}}_i \frac{\partial}{\partial  \boldsymbol{x}_i}   \right)^k \boldsymbol{\lambda}(\boldsymbol{x})  \right|_{x = \bar{\boldsymbol{x}}}
\label{eqn_par_expans}
\end{align}
Using (\ref{eqn_par_expans}), we can evaluate the true mean of (\ref{eqn_ExpectedYtrue}) as
\begin{align}
\bar{\boldsymbol{y}}  = &  \boldsymbol{\lambda}(\bar{\boldsymbol{x}}) +  \left\{     \sum_{i,j = 1}^{n} \frac{\boldsymbol{P}_{ij} }{2!}    \frac{\partial^2 \boldsymbol{\lambda}}{\partial  \boldsymbol{x}_i \partial  \boldsymbol{x}_j}        +    \sum_{i,j,k = 1}^{n} \frac{\boldsymbol{S}_{ijk} }{3!}     \frac{\partial^3 \boldsymbol{\lambda}}{\partial  \boldsymbol{x}_i \partial  \boldsymbol{x}_j \partial  \boldsymbol{x}_k }     \right. \nonumber \\
& \left. +  \sum_{i,j,k,l = 1}^{n}  \frac{\boldsymbol{K}_{ijkl}}{4!}   \frac{\partial^4 \boldsymbol{\lambda}}{\partial  \boldsymbol{x}_i \partial  \boldsymbol{x}_j \partial  \boldsymbol{x}_k \partial  \boldsymbol{x}_l }    \right\}_{ \boldsymbol{x} = \bar{\boldsymbol{x}}} + \cdots  
 \label{eqn_ExpectedYtrue2}
\end{align}  
where $\boldsymbol{P}_{ij} = \mathbb{E}[\tilde{\boldsymbol{x}}_i \tilde{\boldsymbol{x}}_j]$, $\boldsymbol{S}_{ijk} =  \mathbb{E}[\tilde{\boldsymbol{x}}_i \tilde{\boldsymbol{x}}_j \tilde{\boldsymbol{x}}_k]$, and $\boldsymbol{K}_{ijkl} =   \mathbb{E}[\tilde{\boldsymbol{x}}_i \tilde{\boldsymbol{x}}_j \tilde{\boldsymbol{x}}_k \tilde{\boldsymbol{x}}_l]$.

\subsection{True Covariance of the Nonlinear Transformation}
\label{Appendix_covarianceTrue}
The true covariance of $y$ is given as
\begin{align}
\boldsymbol{P}_y & = \mathbb{E} \left[ (\boldsymbol{y} - \bar{\boldsymbol{y}})(\boldsymbol{y} - \bar{\boldsymbol{y}})^T \right] \label{eqn_covY}
\end{align} 
Evaluating the expression $\boldsymbol{y} - \bar{\boldsymbol{y}}$, we write
{\begin{align}
\boldsymbol{y} - \bar{\boldsymbol{y}} & =   D_{\tilde{\boldsymbol{x}}} \boldsymbol{\lambda} + \frac{D_{\tilde{\boldsymbol{x}}}^2 \boldsymbol{\lambda} }{2!}+ \frac{D_{\tilde{\boldsymbol{x}}}^3 \boldsymbol{\lambda} }{3!}   - \mathbb{E}\left[\frac{D_{\tilde{\boldsymbol{x}}}^2 \boldsymbol{\lambda}}{2!} + \frac{D_{\tilde{\boldsymbol{x}}}^3 \boldsymbol{\lambda}}{3!} \right] + \cdots 
 \label{eqn_yYbar}
\end{align}  
Substituting~(\ref{eqn_yYbar}) into (\ref{eqn_covY}) gives
\begin{align}
\boldsymbol{P}_y  & = \mathbb{E} \left[D_{\tilde{\boldsymbol{x}}} \boldsymbol{\lambda} (D_{\tilde{\boldsymbol{x}}} \boldsymbol{\lambda} )^T   +    \frac{D_{\tilde{\boldsymbol{x}}}^2 \boldsymbol{\lambda}  ( D_{\tilde{\boldsymbol{x}}} \boldsymbol{\lambda} )^T}{2!}    +      \frac{D_{\tilde{\boldsymbol{x}}} \boldsymbol{\lambda} (D_{\tilde{\boldsymbol{x}}}^2 \boldsymbol{\lambda})^T }{2!} \right. \nonumber \\
& \quad \left. +   \frac{D_{\tilde{\boldsymbol{x}}}^3 \boldsymbol{\lambda}  ( D_{\tilde{\boldsymbol{x}}} \boldsymbol{\lambda} )^T}{3!}    +      \frac{D_{\tilde{\boldsymbol{x}}} \boldsymbol{\lambda}   (D_{\tilde{\boldsymbol{x}}}^3 \boldsymbol{\lambda})^T }{3!}     + \frac{D_{\tilde{\boldsymbol{x}}}^2 \boldsymbol{\lambda} (D_{\tilde{\boldsymbol{x}}}^2 \boldsymbol{\lambda})^T }{2! \times 2!}   \right] \nonumber \\
& \quad+ \mathbb{E} \left[  \frac{D_{\tilde{\boldsymbol{x}}}^2 \boldsymbol{\lambda} }{2!}    \right] \mathbb{E} \left[  \frac{D_{\tilde{\boldsymbol{x}}}^2 \boldsymbol{\lambda} }{2!}   \right]^T + \cdots
 \label{eqn_Pyt1}
\end{align}  
We note that we can write the first term in the above equation as  
\begin{align}
\mathbb{E} \left[D_{\tilde{\boldsymbol{x}}} \boldsymbol{\lambda} (D_{\tilde{\boldsymbol{x}}} \boldsymbol{\lambda} )^T \right] & =   \left. \frac{\partial \boldsymbol{\lambda}}{\partial \boldsymbol{x}} \right|_{\boldsymbol{x} = \bar{\boldsymbol{x}}} \mathbb{E} \left[ \tilde{\boldsymbol{x}}    \tilde{\boldsymbol{x}}^T \right] \left. \frac{\partial \boldsymbol{\lambda}^T}{\partial \boldsymbol{x}} \right|_{\boldsymbol{x} = \bar{\boldsymbol{x}}}   = \boldsymbol{\lambda} \boldsymbol{P} \boldsymbol{\lambda}^T \label{eqn_PytS1}
\end{align}  
Using (\ref{eqn_par_expans}) and (\ref{eqn_PytS1}), we can rewrite the true covariance matrix of (\ref{eqn_Pyt1}) as
\begin{align}
P_y = & \boldsymbol{\lambda} \boldsymbol{P}  \boldsymbol{\lambda}^T  +  
 \left\{    \sum_{i,j,k  = 1}^{n} \frac{\boldsymbol{S}_{ijk}}{2!}   \left[   \frac{\partial^2 \boldsymbol{\lambda}}{\partial  \boldsymbol{x}_i \partial  \boldsymbol{x}_j}  \frac{\partial \boldsymbol{\lambda}^T}{\partial  \boldsymbol{x}_k}   +    \frac{\partial \boldsymbol{\lambda}}{\partial  \boldsymbol{x}_i}    \frac{\partial^2 \boldsymbol{\lambda}^T}{\partial  \boldsymbol{x}_j \partial  \boldsymbol{x}_k}  \right]  \right.
 \nonumber \\
& +  \sum_{i,j,k,l = 1}^{n}  \boldsymbol{K}_{ijkl}     \left[ \frac{1}{3!} \frac{\partial^3 \boldsymbol{\lambda}}{\partial  \boldsymbol{x}_i \partial  \boldsymbol{x}_j \partial  \boldsymbol{x}_k}  \frac{\partial \boldsymbol{\lambda}^T}{\partial  \boldsymbol{x}_l}  \right.  \nonumber \\
 & \left. +  \frac{1}{3!} \frac{\partial \boldsymbol{\lambda}}{\partial  \boldsymbol{x}_i} \frac{\partial^3 \boldsymbol{\lambda}^T}{\partial  \boldsymbol{x}_j \partial  \boldsymbol{x}_k \partial  \boldsymbol{x}_l} + \frac{1}{4} \frac{\partial^2 \boldsymbol{\lambda}}{\partial  \boldsymbol{x}_i \partial  \boldsymbol{x}_j  }  \frac{\partial^2 \boldsymbol{\lambda}^T}{\partial  \boldsymbol{x}_k \partial  \boldsymbol{x}_l  }   \right]   \nonumber \\
& + 
 \left. \left[     \left. \left.  \sum_{i,j = 1}^{n} \frac{\boldsymbol{P}_{ij}}{2}  \frac{\partial^2 \boldsymbol{\lambda}}{\partial  \boldsymbol{x}_i \partial  \boldsymbol{x}_j}      \right] \right[  \cdots  \right]^T   \right\}_{ \boldsymbol{x} = \bar{\boldsymbol{x}}} + \cdots  
 \label{eqn_Pyt2}
\end{align}  
where we have used the notation $\boldsymbol{x} \boldsymbol{x}^T = \boldsymbol{x} [\cdots]^T$.

\section{Approximation of Means and Covariances using the Generalized Unscented Transform}
We analytically show the accuracy in capturing the true mean and true covariance of $\boldsymbol{y} = \boldsymbol{\lambda}(\boldsymbol{x})$ when using our $2n+1$ sigma points. We also show that our sigma point transformations give improved accuracy by capturing the diagonal components of the skewness and kurtosis tensors. We define $\tilde{\boldsymbol{\chi}}_{[i]} =  \boldsymbol{\chi}_{[i]} -  \bar{\boldsymbol{x}}$ while recalling that $\boldsymbol{C} =  \boldsymbol{P} \boldsymbol{P}^T$. We note that
\begin{align}
\sum_{i = 1}^{2n} \boldsymbol{w}_i D_{\tilde{\boldsymbol{x}}}^k \boldsymbol{\lambda}  & = \sum_{i = 1}^{2n} \boldsymbol{w}_i\left. \left( \sum_{j = 1}^{n}   \tilde{\boldsymbol{x}}_j \frac{\partial}{\partial  \boldsymbol{x}_j}   \right)^k \boldsymbol{\lambda}(\boldsymbol{x})  \right|_{ \boldsymbol{x} = \bar{\boldsymbol{x}}} 
\label{eqn_weightpar_expans}
\end{align}

\subsection{Approximation of the Mean}
\label{Appendix_approximatedMean}
The approximated mean is given as 
\begin{align*}
\hat{\bar{\boldsymbol{y}}} &=     \sum_{i = 0}^{2n} \boldsymbol{w}_i \boldsymbol{\lambda} ( \boldsymbol{\chi}_{[i]}  )   \nonumber \\
& =   \sum_{i = 0}^{2n} \boldsymbol{w}_i  \left[ \boldsymbol{\lambda}(\bar{\boldsymbol{x}}) +   D_{\tilde{\boldsymbol{\chi}}_{[i]}} \boldsymbol{\lambda} + \frac{D_{\tilde{\boldsymbol{\chi}}_{[i]}}^2 \boldsymbol{\lambda}}{2!} + \frac{D_{\tilde{\boldsymbol{\chi}}_{[i]}}^3 \boldsymbol{\lambda}}{3!}  + \cdots  \right]    \nonumber \\
 &=  \boldsymbol{\lambda}(\bar{\boldsymbol{x}}) +   \sum_{i = 1}^{2n}  \boldsymbol{w}_i  \left[   D_{\tilde{\boldsymbol{\chi}}_{[i]}} \boldsymbol{\lambda} +  \frac{D_{\tilde{\boldsymbol{\chi}}_{[i]}}^2 \boldsymbol{\lambda}}{2!} + \frac{D_{\tilde{\boldsymbol{\chi}}_{[i]}}^3 \boldsymbol{\lambda} }{3!} + \cdots   \right]  
\end{align*}
Using (\ref{eqn_weightpar_expans}), we can evaluate the above equation as
\begin{align}
\hat{\bar{\boldsymbol{y}}} =&  \boldsymbol{\lambda}(\bar{\boldsymbol{x}}) + \left\{ \sum_{i,j = 1}^{n}    \frac{\boldsymbol{P}_{ij}}{2}         \frac{\partial \boldsymbol{\lambda} }{\partial  \boldsymbol{x}_i \partial  \boldsymbol{x}_j}       +   \sum_{i,j,k = 1}^{n}   \frac{\hat{\boldsymbol{S}}_{ijk}}{3!}   \frac{\partial^3 \boldsymbol{\lambda}}{\partial  \boldsymbol{x}_i \partial  \boldsymbol{x}_j \partial  \boldsymbol{x}_k}      \right.  \nonumber \\  
& \left. + \sum_{i,j,k,l = 1}^{n}    \frac{\hat{\boldsymbol{K}}_{ijkl} }{4!}        \frac{\partial^4 \boldsymbol{\lambda}}{\partial  \boldsymbol{x}_i \partial  \boldsymbol{x}_j \partial  \boldsymbol{x}_k  \partial  \boldsymbol{x}_l}      \right\}_{ \boldsymbol{x} = \bar{\boldsymbol{x}}}         + \cdots \label{eqn_y_mean_approxi}
\end{align}
where $ \sum_{i = 1}^{2n} \boldsymbol{w}_i  \tilde{\boldsymbol{\chi}}_{ji} \tilde{\boldsymbol{\chi}}_{ki}  = \boldsymbol{P}_{jk}$,  $ \sum_{i = 1}^{2n} \boldsymbol{w}_i  \tilde{\boldsymbol{\chi}}_{ji} \tilde{\boldsymbol{\chi}}_{ki} \tilde{\boldsymbol{\chi}}_{li} = \hat{\boldsymbol{S}}_{jkl}$ and $\sum_{i = 1}^{2n} \boldsymbol{w}_i  \tilde{\boldsymbol{\chi}}_{ji} \tilde{\boldsymbol{\chi}}_{ki} \tilde{\boldsymbol{\chi}}_{li} \tilde{\boldsymbol{\chi}}_{mi} = \hat{\boldsymbol{K}}_{jklm}$.

In the Section \ref{section_sigmaAccuracy}, we already showed that we can accurately capture the diagonal components of the skewness and kurtosis tensors because $\hat{\boldsymbol{S}}_{jkl} = \boldsymbol{S}_{jkl}$ whenever $j = k = l$ and $\hat{\boldsymbol{K}}_{jklm} = \boldsymbol{K}_{jklm}$ whenever $j = k = l = m$. Therefore, by comparing (\ref{eqn_y_mean_approxi}) with the true mean of (\ref{eqn_ExpectedYtrue2}), we can see that our sigma points improves on the accuracy of propagating the mean of a nonlinear transformation.

\subsection{Approximation of the Covariance}
\label{Appendix_covApproximated}
The approximated covariance can be evaluated using the expression
\begin{align}
P_u = \sum_{i = 1}^{2n} \boldsymbol{w}_i \left[ \boldsymbol{\mathcal{Y}}_{[i]} - \hat{\bar{\boldsymbol{y}}} \right]\left[ \boldsymbol{\mathcal{Y}}_{[i]} - \hat{\bar{\boldsymbol{y}}}  \right]^T \label{eqn_Ycov1}
\end{align}
From
\begin{align}
\boldsymbol{\mathcal{Y}}_{[i]} - \hat{\bar{\boldsymbol{y}}}   = &  D_{\tilde{\boldsymbol{\chi}}_{[i]}} \boldsymbol{\lambda} +  \frac{D_{\tilde{\boldsymbol{\chi}}_{[i]}}^2 \boldsymbol{\lambda}}{2!} + \frac{D_{\tilde{\boldsymbol{\chi}}_{[i]}}^3 \boldsymbol{\lambda} }{3!}  \nonumber \\
&  -
    \sum_{j = 1}^{2n}  \boldsymbol{w}_j  \left[ \frac{D_{\boldsymbol{\chi}_{[j]}}^2 \boldsymbol{\lambda}}{2!} + \frac{D_{\boldsymbol{\chi}_{[j]}}^3 \boldsymbol{\lambda}}{3!}  \right] + \cdots  \label{eqn_ybarDiff}
\end{align}
Substituting~(\ref{eqn_ybarDiff}) into (\ref{eqn_Ycov1}) and multiplying out gives 
\begin{align}
P_u  = & \sum_{i = 1}^{2n} \boldsymbol{w}_i \left[ D_{\tilde{\boldsymbol{\chi}}_{[i]}} \boldsymbol{\lambda}  (D_{\tilde{\boldsymbol{\chi}}_{[i]}} \boldsymbol{\lambda})^T +  \frac{D_{\tilde{\boldsymbol{\chi}}_{[i]}}^2 \boldsymbol{\lambda} (D_{\tilde{\boldsymbol{\chi}}_{[i]}} \boldsymbol{\lambda})^T }{2!} \right. \nonumber \\
&  \left. +    \frac{ D_{\tilde{\boldsymbol{\chi}}_{[i]}} \boldsymbol{\lambda} (D_{\tilde{\boldsymbol{\chi}}_{[i]}}^2 \boldsymbol{\lambda})^T}{2!}    +  \frac{D_{\tilde{\boldsymbol{\chi}}_{[i]}}^3 \boldsymbol{\lambda}  (D_{\tilde{\boldsymbol{\chi}}_{[i]}} \boldsymbol{\lambda})^T}{3!}   \right. \nonumber \\
& \left.  +   \frac{ D_{\tilde{\boldsymbol{\chi}}_{[i]}} \boldsymbol{\lambda}  (D_{\tilde{\boldsymbol{\chi}}_{[i]}}^3 \boldsymbol{\lambda})^T}{3!}   + 
\frac{D_{\tilde{\boldsymbol{\chi}}_{[i]}}^2 \boldsymbol{\lambda} (D_{\tilde{\boldsymbol{\chi}}_{[i]}}^2 \boldsymbol{\lambda})^T  }{2! \times 2!}    \right]  \nonumber \\
& 
+ \left. \left.   \left[ \sum_{j = 1}^{2n}  w_j  \frac{D_{\boldsymbol{\chi}_{[j]}}^2 \boldsymbol{\lambda}}{2!}     \right] \right[  \cdots  \right]^T + \cdots \label{eqn_Ycov2}
\end{align}
For the first term in~(\ref{eqn_Ycov2}),  
\begin{align}
\sum_{i = 1}^{2n}  \boldsymbol{w}_i D_{\tilde{\boldsymbol{\chi}}_{[i]}} \boldsymbol{\lambda} (D_{\tilde{\boldsymbol{\chi}}_{[i]}} \boldsymbol{\lambda})^T &=       \sum_{j,k = 1}^{n} \sum_{i = 1}^{2n}  \boldsymbol{w}_i  \tilde{\boldsymbol{\chi}}_{ji}   \tilde{\boldsymbol{\chi}}_{ki} \left. \frac{\partial \boldsymbol{\lambda}  }{\partial  \boldsymbol{x}_j}      \frac{\partial \boldsymbol{\lambda}^T }{\partial  \boldsymbol{x}_k}    \right|_{\boldsymbol{x} = \bar{\boldsymbol{x}}}    \nonumber \\
& = \sum_{j,k = 1}^{n}             \left. \frac{\partial \boldsymbol{\lambda}  }{\partial  \boldsymbol{x}_j}   \right|_{\boldsymbol{x} = \bar{\boldsymbol{x}}}   P_{jk}  \left.   \frac{\partial \boldsymbol{\lambda}^T  }{\partial  \boldsymbol{x}_k}     \right|_{\boldsymbol{x} = \bar{\boldsymbol{x}}}   \nonumber \\
& = \boldsymbol{\lambda} \boldsymbol{P} \boldsymbol{\lambda}^T  \label{eqn_dxP1}
\end{align}
Using (\ref{eqn_weightpar_expans}) and (\ref{eqn_dxP1}), we can rewrite the approximated covariance matrix of (\ref{eqn_Ycov2}) as
\begin{align}
P_u = & \boldsymbol{\lambda} P \boldsymbol{\lambda}^T 
+ \left\{   \sum_{i,j,k = 1}^{n}  \frac{\hat{\boldsymbol{S}}_{ijk}}{2!}     \left[ \frac{\partial^2 \boldsymbol{\lambda}}{\partial  \boldsymbol{x}_i \partial  \boldsymbol{x}_j }  \frac{\partial \boldsymbol{\lambda}^T}{\partial  \boldsymbol{x}_k} + \frac{\partial \boldsymbol{\lambda}}{\partial  \boldsymbol{x}_i }  \frac{\partial^2 \boldsymbol{\lambda}^T}{\partial \boldsymbol{x}_j  \boldsymbol{x}_k}     \right]   \right.
\nonumber \\
& +  \sum_{i,j,k,l = 1}^{n} \hat{\boldsymbol{K}}_{ijkl}   \left[   \frac{1}{3!} \frac{\partial^3 \boldsymbol{\lambda}}{\partial  \boldsymbol{x}_i \partial  \boldsymbol{x}_j  \partial  \boldsymbol{x}_k}  \frac{\partial \boldsymbol{\lambda}^T}{\partial  \boldsymbol{x}_l}  \right.   \nonumber \\  
& \left. + \frac{1}{3!} \frac{\partial \boldsymbol{\lambda}}{\partial  \boldsymbol{x}_i}    \frac{\partial^3 \boldsymbol{\lambda}^T}{\partial  \boldsymbol{x}_j \partial  \boldsymbol{x}_k  \partial  \boldsymbol{x}_l}   +
 \frac{1}{4} \frac{\partial^2 \boldsymbol{\lambda}}{\partial  \boldsymbol{x}_i \partial  \boldsymbol{x}_j}    \frac{\partial^2 \boldsymbol{\lambda}^T}{  \partial  \boldsymbol{x}_k  \partial  \boldsymbol{x}_l}   \right]
\nonumber \\
&  +  \left.  \left. \left.  \left[     \sum_{i,j = 1}^{n}  \frac{\boldsymbol{P}_{ij}}{2}  \frac{\partial^2 \boldsymbol{\lambda}}{\partial  \boldsymbol{x}_i \partial  \boldsymbol{x}_j}   \right] \right[   \cdots \right] ^T  \right\}_{\boldsymbol{x} = \bar{\boldsymbol{x}}}    + \cdots   \label{eqn_YcovApprox}
\end{align}
Comparing (\ref{eqn_YcovApprox}) with the true covariance of (\ref{eqn_Pyt2}), we can see that our sigma points improves on the accuracy of propagating the covariance of a nonlinear transformation because we are able to accurately capture the diagonal components of the skewness and kurtosis tensors.

\bibliographystyle{IEEEtran}
\bibliography{references}

\end{document}